\documentclass[twocolumn, dvipdfmx]{aastex62}

\newcommand{\HeI}{{He}~{\sc i }}

\newcommand{\IRAS}{IRAS 04101+3103}

\accepted{17 Sep 2019}

\usepackage{ulem}
\usepackage{url}
\usepackage{color}

\shorttitle{Mass flow processes around young intermediate-mass stars}

\shortauthors{Yasui et al.}

\begin{document}

\title{Possible progression of mass flow processes around young
intermediate-mass stars based on 
high-resolution near-infrared spectroscopy. I. Taurus}

\correspondingauthor{Chikako Yasui}
\email{ck.yasui@nao.ac.jp}

\author{Chikako Yasui}
\affil{National Astronomical Observatory of Japan, 2-21-1
Osawa, Mitaka, Tokyo 181-8588, Japan, 2-21-1 Osawa, Mitaka, Tokyo
181-8588, Japan} 
\affil{Laboratory of Infrared High-resolution spectroscopy
(LIH), Koyama Astronomical Observatory, Kyoto Sangyo University,
Mo-toyama, Kamigamo, Kita-ku, Kyoto 603-8555, Japan} 

\author{Satoshi Hamano}
\affil{Laboratory of Infrared High-resolution spectroscopy
(LIH), Koyama Astronomical Observatory, Kyoto Sangyo University,
Mo-toyama, Kamigamo, Kita-ku, Kyoto 603-8555, Japan} 

\author{Kei Fukue}
\affil{Laboratory of Infrared High-resolution spectroscopy
(LIH), Koyama Astronomical Observatory, Kyoto Sangyo University,
Mo-toyama, Kamigamo, Kita-ku, Kyoto 603-8555, Japan} 

\author{Sohei Kondo}
\affil{Laboratory of Infrared High-resolution spectroscopy
(LIH), Koyama Astronomical Observatory, Kyoto Sangyo University,
Mo-toyama, Kamigamo, Kita-ku, Kyoto 603-8555, Japan} 

\author{Hiroaki Sameshima}
\affil{Laboratory of Infrared High-resolution spectroscopy
(LIH), Koyama Astronomical Observatory, Kyoto Sangyo University,
Mo-toyama, Kamigamo, Kita-ku, Kyoto 603-8555, Japan}

\author{Keiichi Takenaka}
\affil{Laboratory of Infrared High-resolution spectroscopy
(LIH), Koyama Astronomical Observatory, Kyoto Sangyo University,
Mo-toyama, Kamigamo, Kita-ku, Kyoto 603-8555, Japan} 
\affil{Department of Astrophysics and Atmospheric Sciences,
Faculty of Sciences, Kyoto Sangyo University, Motoyama, Kamigamo,
Kita-ku, Kyoto 603-8555, Japan}

\author{Noriyuki Matsunaga} 
\affil{Department of Astronomy, Graduate School of Science, The
University of Tokyo, 7-3-1 Hongo, Bunkyo-ku, Tokyo 113-0033, Japan}
\affil{Laboratory of Infrared High-resolution spectroscopy
(LIH), Koyama Astronomical Observatory, Kyoto Sangyo University,
Mo-toyama, Kamigamo, Kita-ku, Kyoto 603-8555, Japan} 

\author{Yuji Ikeda} 
\affil{Laboratory of Infrared High-resolution spectroscopy
(LIH), Koyama Astronomical Observatory, Kyoto Sangyo University,
Mo-toyama, Kamigamo, Kita-ku, Kyoto 603-8555, Japan} 
\affil{Photocoding, 460-102 Iwakura-Nakamachi, Sakyo-ku,
Kyoto,606-0025, Japan} 

\author{Hideyo Kawakita} 
\affil{Laboratory of Infrared High-resolution spectroscopy
(LIH), Koyama Astronomical Observatory, Kyoto Sangyo University,
Mo-toyama, Kamigamo, Kita-ku, Kyoto 603-8555, Japan} 
\affil{Department of Astrophysics and Atmospheric Sciences,
Faculty of Sciences, Kyoto Sangyo University, Motoyama, Kamigamo,
Kita-ku, Kyoto 603-8555, Japan} 

\author{Shogo Otsubo} 
\affil{Laboratory of Infrared High-resolution spectroscopy
(LIH), Koyama Astronomical Observatory, Kyoto Sangyo University,
Mo-toyama, Kamigamo, Kita-ku, Kyoto 603-8555, Japan} 
\affil{Department of Astrophysics and Atmospheric Sciences,
Faculty of Sciences, Kyoto Sangyo University, Motoyama, Kamigamo,
Kita-ku, Kyoto 603-8555, Japan} 

\author{Ayaka Watase} 
\affil{Laboratory of Infrared High-resolution spectroscopy
(LIH), Koyama Astronomical Observatory, Kyoto Sangyo University,
Mo-toyama, Kamigamo, Kita-ku, Kyoto 603-8555, Japan} 
\affil{Department of Astrophysics and Atmospheric Sciences,
Faculty of Sciences, Kyoto Sangyo University, Motoyama, Kamigamo,
Kita-ku, Kyoto 603-8555, Japan}

\author{Daisuke Taniguchi} 
\affil{Department of Astronomy, Graduate School of Science, The
University of Tokyo, 7-3-1 Hongo, Bunkyo-ku, Tokyo 113-0033, Japan}

\author{Misaki Mizumoto} 
\affil{Centre for Extragalactic Astronomy, Department of
Physics, University of Durham, South Road, Durham DH1 3LE, UK}

\author{Natsuko Izumi} 
\affil{National Astronomical Observatory of Japan, 2-21-1
Osawa, Mitaka, Tokyo 181-8588, Japan, 2-21-1 Osawa, Mitaka, Tokyo
181-8588, Japan} 
\affil{Laboratory of Infrared High-resolution spectroscopy
(LIH), Koyama Astronomical Observatory, Kyoto Sangyo University,
Mo-toyama, Kamigamo, Kita-ku, Kyoto 603-8555, Japan} 

\author{Naoto Kobayashi} 
\affil{Institute of Astronomy, University of Tokyo, 2-21-1
Osawa, Mitaka, Tokyo 181-0015, Japan}
\affil{Kiso Observatory, Institute of Astronomy, School of
Science, The University of Tokyo, 10762-30, Mitake, Kiso-machi,
Kiso-gun, Nagano 397-0101, Japan}
\affil{Laboratory of Infrared High-resolution spectroscopy
(LIH), Koyama Astronomical Observatory, Kyoto Sangyo University,
Mo-toyama, Kamigamo, Kita-ku, Kyoto 603-8555, Japan} 



\begin{abstract}



We used the WINERED spectrograph to perform near-infrared
high-resolution spectroscopy (resolving power $R = 28$,000) of 13 young
intermediate-mass stars in the Taurus star-forming region. Based on the
presence of near- and mid-infrared continuum emission, young
intermediate-mass stars can be classified into three different
evolutionary stages: Phases I, II, and III in the order of evolution.
Our obtained spectra ($\lambda = 0.91$--1.35\,$\mu$m) depict \HeI
$\lambda$10830 and P$\beta$ lines that are sensitive to magnetospheric
accretion and winds.
We also investigate five sources each for P$\beta$ and \ion{He}{1}
lines that were obtained from previous studies along with our targets.
We observe that the P$\beta$ profile morphologies in Phases I and II
corresponded to an extensive variety of emission features; however,
these features are not detected in Phase III.
We also observe that the He I profile morphologies are mostly broad
subcontinuum absorption lines in Phase I, narrow subcontinuum absorption
lines in Phase II, and centered subcontinuum absorption features in
Phase III.
Our results indicate that the profile morphologies exhibit a progression
of the dominant mass flow processes: stellar wind and probably
magnetospheric accretion in the very early stage,
magnetospheric accretion and disk wind in the subsequent stage, and no
activities in the final stage.
These interpretations further suggest that opacity in protoplanetary disks
plays an important role in mass flow processes.
Results also indicate that \ion{He}{1} absorption features in
Phase III sources, associated with chromospheric activities
even in such young phases, 
are characteristics of intermediate-mass stars.

\end{abstract}

\keywords{accretion, 
planetary systems: protoplanetary disks --- 
stars: formation --- 
stars: pre-main-sequence ---
stars: winds, outflows ---
infrared: stars}



\section{INTRODUCTION} \label{sec:intro}

Stars are formed owing to the gravitationally induced collapse of cold
molecular gas. While the molecular clouds are contracted during the
collapse, the initial rotation of the star-forming cloud is enormously
magnified by the conservation of angular momentum. In this process, the
rotating circumstellar disks are formed, and majority of the material of
a typical star is accreted through its disk \citep{Hartmann2009book}.
Although almost all of the angular momentum ($>$99\,\%) must be removed
to form an observable star, we lack a conclusive theory explaining the 
mechanics of the angular momentum transfer (``angular momentum
problem'', \citealt{Bodenheimer1995}).
As one of the most important processes, this study focuses on the mass
flow processes occurring very close (radius $r \lesssim 1$\,AU) to
central stars. In this region, a large population of close-in planetary
systems has been reported, even around young stellar objects (YSOs)
younger than 10\,Myr (e.g., \citealt{Donati2016}).
However, the manner in which planets are formed is still being debated
\citep{Winn2015ARAA}.
Because the interaction between the stars and disks at which planets are
formed occurs mainly in the inner disk region, it is essential to
investigate the conditions of this region from the planet formation
viewpoint.

As for mass flow processes, magnetospheric accretion and outflowing gas
have been proposed.
Because direct imaging for separations of less than 1\,AU is considered
to be very difficult using the existing instruments, spectroscopy has
been employed. Traditionally, H$\alpha$, NaD, \ion{Ca}{2} H\&K,
\ion{Mg}{2} h\&k lines have been used for conducting diagnostics
\citep[e.g.,][]{{Mundt1984},{Calvet1992}}.
For mass accretion, these diagnostics have permitted the development of
a general picture with respect to various factors, including the mass
accretion rate, evolution, geometry, {and} stellar mass
dependence \citep[e.g.,][]{{Muzerolle1998b}, {Hartmann1998},
{Muzerolle2004}, {Calvet2004}}.
However, the actual mechanics of outflowing gas have not yet been
completely elucidated. There are at least three possible configurations
of wind formation: a disk wind \citep{Ustyugova1995}, an X-wind
\citep{Shu1994}, and a stellar wind \citep{Hartmann1982}.
To distinguish between different theoretical models, constraints on the
nature or location of the wind-launching region are very important.
However, the aforementioned traditional spectral lines exhibit
blueshifted absorption superposed on the broad emission features.
This observation signifies the presence of high-velocity wind close to
the star \citep{Ardila2002}, yielding little information on the nature
or location of the wind-launching region.

Recently, the \ion{He}{1} 10,830\,\AA\ line has been suggested as a
particularly suitable and sensitive probe of both mass accretion and
outflow \citep{Edwards2006}.
\citet{Edwards2006} exhibited that $\sim$70\,\% of their 39 sample
classical T-Tauri stars (CTTSs) exhibit a blueshifted subcontinuum
absorption that can be assumed to be a probe of outflow.
This is in contrast to H$\alpha$, for which only $\sim$10\,\% of the
stars show a similar type of absorption component.
Additionally, $\sim$50\,\% of their CTTSs show redshifted absorption
below the continuum in \ion{He}{1}; this can be considered to be a probe
of mass accretion. 
They also highlighted that the direct probes of the outflowing gas
launch region can be created only by high-resolution spectroscopy of the
\ion{He}{1} line. While the targets for \citet{Edwards2006} were
primarily low-mass stars, \citet{Oudmaijer2011} and \citet{Cauley2014}
presented a \ion{He}{1} $\lambda$10830 profile for Herbig Ae/Be stars
(HAeBes), which are intermediate-mass pre-main sequence stars.
The calculated \ion{He}{1} model profile (e.g., \citealt{Kwan2007},
\citealt{Fischer2008}, \citealt{Kurosawa2011}, \citealt{Kurosawa2012})
suggests that the mass flow processes occurring in the inner part of the
protoplanetary disks (disk wind, stellar wind, and magnetospheric
accretion) can be distinguished based on the \ion{He}{1} profile
morphologies.
Further, the hydrogen Paschen series can also serve as diagnostics for
magnetospheric accretion. Because these series lines are of lower
opacity when compared to the H$\alpha$ lines, they depict signatures of
infall of accreting material in a magnetospheric funnel flow more
clearly, whereas H$\alpha$ often do not show such signatures
\citep{Muzerolle2001}.
\citet{{Folha1999},{Folha2001}} presented the P$\beta$ line profiles for
49 T Tauri stars and observed that inverse P Cygni profile features,
indicating the magnetospheric mass accretion, can be observed at a 
relatively high frequency even though the blueshifted absorption
features are absent.
\citet{Edwards2006} also presented the P$\gamma$ line profiles for 48 T
Tauri stars. The calculated model profiles of the Paschen lines (e.g.,
\citealt{Kurosawa2011}, \citealt{Kurosawa2012}) suggest that the
existence or absence of magnetospheric accretion can be determined based
on the line morphology; however, these lines cannot serve as diagnostics
for winds.

The progression of the dominant mass flow processes through stellar
evolution should also impose constraints on the mechanistic theories.
\citet{Edwards2006} proposed a progression involving veilings, which are
considered to be correlated with the mass accretion rate (e.g.,
\citealt{Beristain2001}).
In particular, these authors suggested that the emission equivalent
width of P$\gamma$ decreases with decreasing veiling. Based on
\ion{He}{1} profiles, \citet{Kwan2007} used a comparison of their
theoretical line profiles with the profiles observed by
\citet{Edwards2006} to suggest that the wind signatures of stars with
high disk accretion rates are more likely to indicate stellar wind than
disk wind.
Further, although a moderate trend is evident in these factors, this
pattern does not seem to be valid for all the stellar sources. This
inconsistency may be attributed to an extensive variety of disk
properties that have been revealed by recent observational studies,
e.g., mass accretion rate, disk lifetime, and disk mass.
The median disk lifetime has been estimated to be approximately 3\,Myr 
even though the lifetimes vary from less than 1\,Myr to a maximum of
10\,Myr. The mass accretion rate and disk mass also exhibit large
dispersions, $\sim$$\pm$1\,dex (for the mass accretion rate, see Figs.~2
and 3 in \citealt{Hartmann2008}; for the disk mass, see Fig.~5 in
\citealt{Williams2011} and Fig.~7 in \citealt{Ansdell2017}).
Furthermore, because the entirety of dust and gas in the inner and outer
disks is dispersed almost simultaneously; at 0.5\,Myr
\citep{Williams2011}, there appear to be no clear indicators of the
protoplanetary disk evolutionary stages.
However, this picture is obtained from low-mass stars
($\lesssim$1\,$M_\odot$), which constitute the majority in the solar
neighborhood. Recent observations suggest that the protoplanetary disks
surrounding the intermediate-mass stars show different evolutionary
stages; for the intermediate-mass stars ($\gtrsim$1.5\,$M_\odot$), only
the innermost disk regions that are traced with the near-infrared (NIR)
 {\it K}-band
($r\sim 0.3$\,AU) disperse at a very early time as compared to the outer
disks by $\sim$3 to 4\,Myr \citep{Yasui2014}.
Therefore, based on the presence of innermost and outer disks, the
intermediate-mass stars can be categorized into three phases; further,
the intermediate-mass stars should be appropriate targets for
investigating the time variation of dominant mass flow processes.
Based on the \ion{He}{1} line profiles, \citet{Cauley2014} suggested
that HAe stars exhibit signatures of magnetospheric accretion and
stellar wind even though they do not exhibit disk wind.
Although these authors have suggested that these processes are
characteristic of the intermediate-mass stars, this may only be valid
for the Phase I sources to which the HAe stars are comparable. To
bridge this gap, it is necessary to observe young intermediate-mass
stars in all the evolutionary phases.

In this study, we present the results of a census of \ion{He}{1}
$\lambda$10830 and P$\beta$ in 13 young intermediate-mass stars in the
Taurus star-forming region, which is one of the nearest active
star-forming regions (distance of $D=140$\,pc;
\citealt{Kenyon2008book}). 
Given its proximity, this region can be investigated using X-ray to
radio wavelengths, and the members of Taurus have been more
comprehensively identified (from early type stars to brown dwarfs) than 
those of any other star-forming region. Given the young age of the
Taurus complex ($\sim$1.5\,Myr \citep{Barrado2003}),
most members are still surrounded by circumstellar materials. Therefore,
Taurus is the best target for conducting our first star-formation study
with WINERED (Warm INfrared Echelle spectrograph for REalizing Decent
infrared high-resolution spectroscopy). 
WINERED is a spectrograph of high sensitivity and high
resolution (with two resolution modes of
$R=28$,000 and $R=70$,000), specifically customized for short NIR bands
at 0.9 to 1.35\,$\mu$m \citep{Yasui2008}. WINERED received its first
light in 2012 May at the Araki 1.3\,m telescope and is now attached to
the New Technology Telescope (NTT) \citep{Ikeda2018}; further, it is
planned to attach WINERED to the Magellan Telescope.
We are continuing to investigate YSOs in the target star-forming
regions, including Tr 37 and NGC 7160.
By considering that the lifetime of the protoplanetary disks is
approximately 10\,Myr (e.g., \citealt{Hernandez2008}), these regions are
considered to be in different evolutionary phases than Taurus. Tr 37
($\sim$4\,Myr) is in the middle phase, while NGC 7160 ($\sim$10\,Myr) is
in the final phase. Our results for the other regions will be provided
in a future work.

This study can be organized as follows. In Section~2, we discuss the
target selection, observations, and data reduction.  In Section~3, we
present the obtained spectra of the Paschen series and \ion{He}{1}
$\lambda$10830 and determine the excess continuum emission and residual
profiles. We further discuss the properties of the P$\beta$ and
\ion{He}{1} $\lambda$10830 lines.
In Section~4, we discuss the statistics of the line profile morphologies
and present an interpretation of the profile morphology of P$\beta$ and
\ion{He}{1} $\lambda$10830 in different evolutionary phases of the
protoplanetary disks in combination with other intermediate-mass stars
that have been observed in previous studies.
We further discuss the possible progression of dominant mass flow
processes in the evolution of protoplanetary disks. After comparing the
results with those of the previous studies, we finally discuss the
implications for theories of mass flow processes.


\section{TARGET SELECTION, OBSERVATION AND DATA REDUCTION}

\subsection{Selection and evolutionary phase of targets}
\label{sec:targets}

We selected target intermediate-mass stars in the Taurus star-forming
region. We defined stars within the mass range of 1.5--7\,$M_\odot$ as
intermediate-mass stars \citep{Yasui2014}.
The intermediate-mass stars were selected in a similar manner to that of
\citet{Yasui2014}; assuming a typical age of the Taurus star-forming
region, $\sim$1.5\,Myr \citep{Barrado2003}, we obtained the spectral
types of cluster members from the literature.
For the lower mass limit, we used a spectral type of K5 corresponding
to 1.5\,$M_\odot$. 
Although \citet{Yasui2014} set the upper limit mass at 7\,$M_\odot$
(spectral type B3), even the most massive stars in Taurus possess masses
of $\sim$4\,$M_\odot$ (spectral type B9), which is substantially lower
than the upper mass limit. Among the candidates selected above, the
final targets are limited to stars with {\it J}-band magnitudes of less
than 9 mag in the Two Micron All Sky Survey (2MASS) Point Source Catalog
(PSC; \citealt{Skrutskie2006}). 
This criterion is required by the limited sensitivity of the Araki
1.3\,m telescope's implementation of WINERED, with which this study's
spectra can be obtained (see Section~\ref{sec:obs}). Therefore,
13 target intermediate-mass stars were selected.
The properties of the selected target stars are summarized in
Table~\ref{tab:target}.

\begin{figure*}[h]
\includegraphics[scale=0.5]{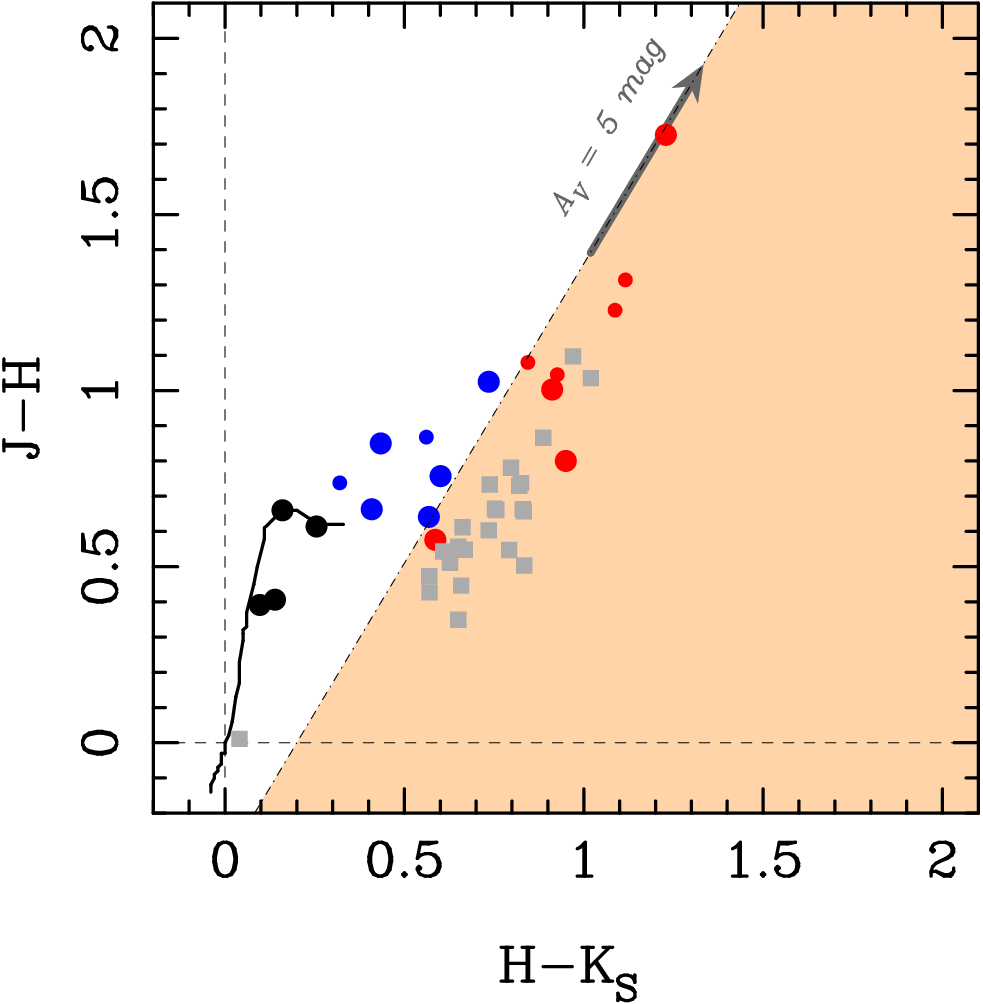}
\caption{$H-K_S$ vs. $J-H$ color--color diagram.
The dwarf track \citep{Bessell1988} is shown with a black line, 
while the reddening vector 
\citep{Rieke1985} for $A_V = 5$\,mag is shown with the gray arrow.
The border line, which is parallel to the reddening vector and
distinguishes HAeBe stars from other objects, is shown with a dot-dashed
line.
The region to the right of the border line (orange color) is defined as
the ``excess region'' for intermediate-mass stars \citep{Yasui2014}.
Target intermediate-mass stars in this paper are shown with filled circles: 
Phase I, II, and III sources are shown with red, blue, and black,
respectively.
Six {additional} intermediate-mass stars in the sample{s} of
\citet{Folha2001} and \citet{Edwards2006} are denoted by small symbols
with the same colors as the targets {observed in the present study}.
HAe stars in a sample of \cite{Cauley2014} are shown with gray filled
squares.} \label{fig:JHKcc}
\end{figure*}

In a previous study \citep{Yasui2014}, we derived the intermediate-mass
disk fraction in the near-infrared {\it JHK} photometric bands and in
the mid-infrared (MIR) bands for young clusters in the age range of 0 to
$\sim$10\,Myr.
\citet{Yasui2014} have defined the region to the right of the border
line\footnote{The border line (the dot-dashed line) passes through the
point of $(H-K_S, J-H) = (0.2, 0)$ and is parallel to the reddening
vector (shown with gray arrow in Fig.~\ref{fig:JHKcc}).}  (region
highlighted in orange region in Fig.~\ref{fig:JHKcc}) as the disk excess
region on $H-K$ vs. $J-H$ color---color diagram (Fig.~\ref{fig:JHKcc});
stars located on the disk excess region are recognized as stars traced
by NIR excess emissions.
MIR disk fraction is derived using {\it Spitzer} IRAC [3.6]--[8]
photometric bands. The stars that have SED slope with the MIR bands
($\alpha = d \ln \lambda F \lambda / d \ln \lambda$) of $>$$-$2.2 are
recognized as stars traced by MIR excess emissions \citep{Kennedy2009}.
In the case of 2.5\,$M_\odot$-stars, NIR and MIR continuum emissions
trace innermost dust disk ($r \sim 0.3$\,AU) and outer dust disk ($r
\gtrsim 5$\,AU), respectively.
\citet{Yasui2014} have suggested that in the very early phase, the 
intermediate-mass stars disperse only their innermost dust disks.
They further disperse their outer dust disks with a time lag of $\sim$3
to 4\,Myr. Therefore, the presence of innermost disks and outer disks
surrounding the intermediate-mass stars can be a clear evolutionary
indicator. Hereafter, we will refer to the three evolutionary phases of
 ``Phase I'' that contains stars with both innermost and outer disks,
 ``Phase II'' that contains stars without innermost disks but with outer
 disks, and ``Phase III'' that contains stars without innermost and
 outer disks.

The photometric data for 13 target intermediate-mass stars selected
above are shown in Table~\ref{tab:target}.
The data of NIR bands are obtained from the 2MASS PSC for targets, whose
all {\it JHK} band photometric quality flags are denoted as ``A''
(signal-to-noise $\le$10 for all {\it JHK} bands).
Because AB Aur does not satisfy the condition, we used NIR data reported
by \citet{Kenyon1995}.
We plotted our target sources on $H-K$ vs. $J-H$ diagram in
Fig.~\ref{fig:JHKcc}.
Only four sources (V892 Tau, AB Aur, IRAS 04101+3103, and T Tau) are
located in the disk excess region, suggesting that  they have innermost 
dust disks.
For MIR data, because IRAC photometric data for most target stars could
not be obtained from the IRAC observations of the Taurus star-forming
region (e.g., \citealt{Hartmann2005}), we adopted $\alpha$ values
derived using Spitzer IRS within $\lambda = 6$--13\,$\mu$m.
Nine sources (V892 Tau, AB Aur, IRAS 04101+3103,
RY Tau, SU Aur, T Tau, RW Aur, V773 Tau, and UX Tau) have 
$\alpha$ values of $\ge$$-$2.2, suggesting that they have 
outer dust disks. 
Therefore, the selected targets cover all the evolutionary 
phases of the protoplanetary disks, with four Phase I sources, five
Phase II sources, and four Phase III sources. 
Four sources (V892 Tau, AB Aur, IRAS 04101+3103, and T Tau) are
categorized into Phase I, five sources (RY Tau, SU Aur, T Tau, RW Aur,
V773 Tau, and UX Tau) are categorized into Phase II, and four sources
(HP Tau/G2, HD 283572, HBC 388, and V410 Tau) are categorized into Phase
III.
Table~\ref{tab:disk_prop} summarizes basic disk properties for all the
targets, showing classification, mass accretion rate, and disk mass.
Notably, no clear distinctive properties are observed  
between Phase I and II sources except for aforementioned 
NIR {\it K}-band excesses.

We indicate that disks around the intermediate-mass stars necessarily
evolve from Phase I to Phase II and, subsequently, to Phase III, while
the disks around the low-mass stars evolve from Phase I to Phase III,
spending a considerably short time in Phase II \citep{Yasui2014}.
It should be noted that HAeBes are categorized into Phase I sources
because they are confirmed to contain NIR {\it K}-band excesses.
Fig.~\ref{fig:JHKcc} plots HAe stars in a sample used by
\citet{Cauley2014}, whose mass range is roughly comparable to that of
our target intermediate-mass stars and whose \ion{He}{1} profile
morphologies will be later compared with those of our targets
(Sections~\ref{sec:HeI}, \ref{sec:HeI}, \ref{sec:Implication}); the
stars are shown with gray squares.
We used NIR data from 2MASS PSC and rejected four sources (V1578 Cyg, HD
17081, HD 37490, and Z CMa), which do not have an ``A'' photometric
quality flag for at least one band in the catalog. 
Thus, we confirmed that all sources, except one source, are located in
the disk excess region having large color excesses.
The exception is HD 141569 ($(J-H, H-K) = (0.01, 0.04)$), 
which \citet{Cauley2014} categorized into potentially misclassified
objects and did not include in their final discussion by indicating 
that it may not be a HAe star.
Fig.~\ref{fig:JHKcc} also shows the intermediate-mass stars in samples
reported by \citet{Folha2001} and \citet{Edwards2006} that satisfy the
conditions for selecting intermediate-mass stars in Taurus, but are
excluded from the sample {used in this study} due to the limited
instrumental sensitivities.
They are six sources and their properties are summarized in
Table~\ref{tab:target_others}.
Four of these sources (CW Tau, GM Aur, DR Tau, and DS Tau) are included
in both the previous studies, while HP Tau and HN Tau are included in
only one of these references
(\citealt{Folha2001} and \citealt{Edwards2006}). 
CW Tau, HP Tau, DR Tau, and HN Tau are categorized into Phase I based on
their NIR and MIR excesses, while GM Aur and DS Tau are categorized into
Phase II.
For reference, CTTSs and weak T-Tauri stars (WTTSs) are almost
comparable to Phase I and Phase III, respectively, in case of low-mass
stars \citep{Yasui2014}.

\subsection{WINERED spectroscopy and data reduction}
\label{sec:obs} 

We obtained the spectra of selected targets using the near-infrared
(NIR) spectrograph WINERED \citep{{Kondo2015},{Ikeda2016}} attached to
the Araki 1.3\,m-telescope at Koyama Astronomical Observatory,
Kyoto-Sangyo University, Japan \citep{Yoshikawa2012}.
WINERED has a 1.7\,$\mu$m cutoff $2048 \times 2048$ HAWAII-2RG infrared
array with a pixel scale of $0''.8$\,pixel$^{-1}$, simultaneously
covering the wavelength range of 0.91--1.35\,$\mu$m.
We used a  1.65$''$-width slit, corresponding to 2\,pixels, which
proves a spectral resolving power of $R=28,300$ or 11\,km\,s$^{-1}$. 
Observations were performed during the following three periods: February
to March 2013; November 2013 to January 2014; and September to October
2014.
Seeing was typically $\sim$5$''$. 
To avoid saturation in the emission lines, the exposure time for each
frame was set to a duration ranging from 300 to 1200\,s.
Four to twelve sets of data were obtained for each target, resulting in
a total exposure time of 1200 to 7200\,s for each target.
As the telluric standard stars, the spectra of bright stars with
spectral types of late B to early A (B9--A2) at similar airmass ($\Delta
\sec z \lesssim 0.2$) were obtained in a similar fashion within 2 hours
of target data acquisition on the same night.  As the sole exception,
the spectra of an M1III spectral type star, HD 1041, as the telluric
standard star were obtained for HP Tau/G2.
We summarize the details of the observation in Table~\ref{tab:obs}.

All the data were reduced by following the standard procedures using
IRAF, including nodding sky subtraction, dome flat-fielding, and
aperture extraction.
The argon lamp spectra that were obtained at the end of the observing
night were used for performing vacuum wavelength calibration.  Each
target spectrum was divided by the spectrum of the standard star to
rectify for atmospheric absorption and instrumental response after the
photospheric features of the standard spectrum were eliminated
\citep{Sameshima2018}.
For each echelle order, the spectra were normalized to 1.0. 
For the continuum level at $\lambda \sim 12700$\,\AA, 
where there are no strong photospheric features in synthetic spectra (see 
Section~\ref{sec:veil_PsF})
for all the targets, the estimated signal to noise ratios (S/N) were
$\sim$30 to 110 (see Table~\ref{tab:obs} for each target). 
The spectra around P$\delta$, P$\gamma$, P$\beta$, and \ion{He}{1}
(orders of $m=56, 51, 44$, and 52, respectively),
reduced by following the above procedure, are depicted for all the
targets with respect to the stellar velocities in
Figure~\ref{Fig:3H_HeI}.
The spectra are sorted by spectral type from the early to late type.

\section{Results} \label{sec:results}

The obtained spectra depict several lines, including those of hydrogen
(Paschen series), \ion{He}{1}, and metals (e.g., \ion{Fe}{1},
\ion{Si}{1}, \ion{Mg}{1}).
We focus on hydrogen lines and \ion{He}{1} $\lambda$10830 in this study
because these are very sensitive to magnetospheric mass accretion and
winds.

\subsection{Hydrogen Paschen series} \label{sec:Paschen}

\begin{figure*}[h]
\includegraphics[scale=1.2]{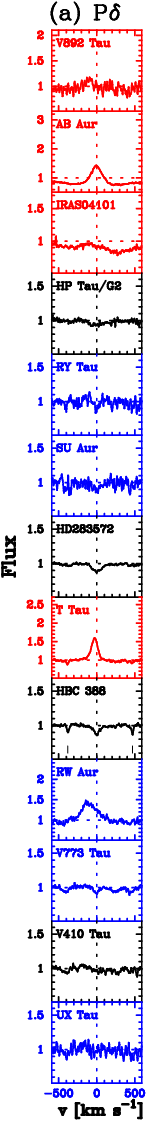}
\hspace{1em}
\includegraphics[scale=1.2]{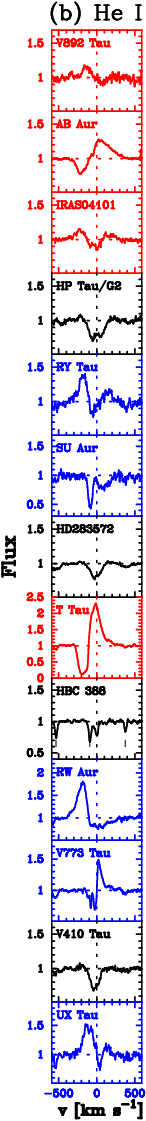}
\hspace{1em}
\includegraphics[scale=1.2]{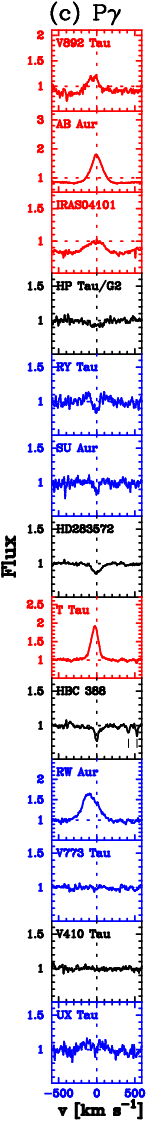}
\hspace{1em}
\includegraphics[scale=1.2]{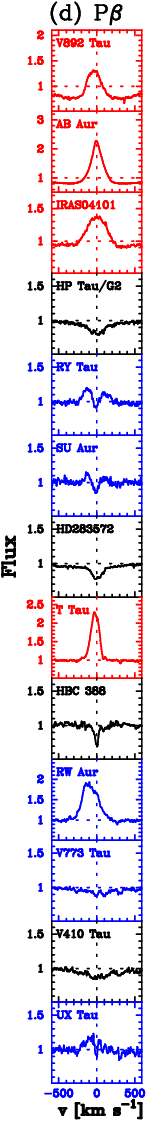}

\caption{Spectral regions of P$\delta$ (a), \ion{He}{1} $\lambda$10830
 (b), P$\gamma$ (c), and P$\beta$ (d).  
Fluxes are normalized to the continuum. 
Velocities are relative to the stellar rest velocities. 
The spectra are sorted by spectral type, from early to late type. 
Spectra for Phase I, II, and III sources are shown with red, blue, and
black, respectively. 
Photospheric features are marked with vertical lines in the panels for
 HBC 388.
Note that the plot range has been widened for some objects with strong
 features.} \label{Fig:3H_HeI}
\end{figure*}

The WINERED wavelength range ($\lambda=0.91$--1.35\,$\mu$m) covers four
hydrogen lines: P$\epsilon$ ($\lambda=9549$\,\AA), P$\delta$
($\lambda=10052$\,\AA), P$\gamma$ ($\lambda=10941$\,\AA), and P$\beta$
($\lambda=12822$\,\AA), where the numbers in parentheses refer to the
vacuum wavelengths.
The spectra of P$\delta$, P$\gamma$, and P$\beta$ are depicted in
Figure~\ref{Fig:3H_HeI} (panels a, b, and d, respectively).
Although each spectrum appears to show similar features across various
targets, the S/Ns of the spectra are higher at longer wavelengths,
resulting in the features of those spectra becoming more prominent.
These tendencies are clearly observed in, e.g., RY Tau, SU Aur, UX Tau,
and HP Tau/G2. Consider UX Tau as an example.  Although the P$\gamma$
spectrum shown in \citet{Edwards2006} exhibits no features, the
Pa$\beta$ spectrum in Figure~\ref{Fig:3H_HeI} depicts significant
features, including double peaks with centered subcontinuum absorption,
thereby making the diagnostics of magnetospheric accretion possible.
Therefore, we focus on the spectra of P$\beta$ among four lines in the
hydrogen Paschen series in this study. This approach can help us to
detect the possible characteristics of profile morphologies in different
evolutionary phases.

\subsection{Removal of photospheric features: 
Estimation of line broadening, radial velocities, and veilings} 
\label{sec:veil_PsF}

The observed spectra comprise circumstellar and photospheric
features. To extract the circumstellar features, it is necessary to
eliminate the photospheric features. The photospheric features are
broadened (line broadening, $v_{\rm broad}$), and shifted because of
radial velocities (RVs); further, they also exhibit continuum
excesses. The excess can be referred to as ``veiling,'' which is defined
as the ratio of the excess emission to the underlying stellar
photosphere (see \citealt{Hartigan1989}). We eliminate the photospheric
features using the following procedure. First, by comparing the observed
spectra with the synthetic spectra, the stellar properties, line
broadenings, radial velocities RVs, and veilings can be obtained. We
further obtain the circumstellar features by subtracting the synthetic
spectra from the observed spectra by considering the aforementioned
stellar properties.

For each object, we constructed the synthetic spectra using the analysis
program SPTOOL
\citep{{Takeda1995},{Takeda2005}}, which is based on the ATLAS9 programs
from \citet{Kurucz1993}. We adopted the metal line list in
\citet{Melendez1999} and the molecular lines from the Vienna Atomic Line
Database (VALD3; \citealt{Ryabchikova2015}). Further, we assumed solar
abundance and spectral type from the literature (see
Table~\ref{tab:target}).
The adopted parameters for the synthetic spectra, effective temperature
($T_{\rm eff}$), surface gravity ($\log g$), and microturbulent velocity
are provided below. The effective temperatures were adopted from
\citet{Gray2005} (Table~B.1. in Appendix~B).
 In a case that the spectral types for targets were not listed in
 \citet{Gray2005}, close spectral types were used
 (see Table~\ref{tab:prop_W}, column 2).
The $\log g$ was calculated with $T_{\rm eff}$, stellar mass ($M_*$),
and stellar luminosity ($L_*$). Because the model dependence of $\log g$
is considered to be very small \citep{Takagi2011}, we used the values of
$T_{\rm eff}$, $M_*$, and $L_*$ from a PMS isochrone model,
\citet{Siess2000}, by considering the age of Taurus (1.5\,Myr old). The
obtained $\log g$ values are 4.1 for spectral type B8, 3.5 for spectral
type K0, and 3.7 for spectral types K3--K5. We adopted the values of 4.1
for late B to early A type stars (V892 Tau, AB Aur, and IRAS
04101+3103), 3.5 for G0--K1 type stars (HP Tau/G2, RY Tau, SU Aur, HD
283572, T Tau, and HBC 388), and 3.7 for K3--K5 type stars (RW Aur, V773
Tau, V410 Tau, and UX Tau). For microturbulent velocity, we adopted
2\,km\,s$^{-1}$ according to \citet{Gray2001}, who observed a tight
correlation between microturbulence and $\log g$ in the spectral range
of late A- to early G-type (see Fig.~7 in \citealt{Gray2001}).

We derived line broadening $v_{\rm broad}$, radial velocity RV, and {\it
Y}- and {\it J}-band veilings ($\gamma_Y$ and $\gamma_J$, respectively)
for each target by fitting the obtained synthetic spectra with the
observed spectra.
First, we roughly estimated $v_{\rm broad}$\footnote{Here, $v_{\rm
broad}$ is parameterized by FWHM. However, because the output velocity
in MPFIT is a convolution of $v_{\rm broad}$ and the instrumental
broadening, and is parameterized by the {\it e}-folding half width, it
is converted to $v_{\rm broad}$ according to \citet{Takeda2008}.}
and RV by fitting the obtained synthetic spectra with the observed
spectra using the automatic profile-fitting program MPFIT
\citep{Takeda1995} included in SPTOOL. 
Note that the veilings are not yet considered at this stage. Among the
spectra shown in Fig.\ref{Fig:3H_HeI} ($m=56$, 52, 51, and 44), our
fitting procedure used spectra with interference order $m=52$, which
included \ion{He}{1} $\lambda$10830, because these spectra show a large
number of relatively deep lines, with normalized counts of less than 0.7
in normalized synthetic spectra at $v_{\rm broad} = 0$\,km\,s$^{-1}$.
The deep lines used for the fitting, in the wavelength ranges of
$\lambda=10710$--10815\,\AA\ and 10855--10875\,\AA, are not affected by
the \ion{He}{1} $\lambda$10830 features and do not include data with
very low S/N due to the edges of the spectra. The lines used for the
stars with spectral types G--K2 were
\ion{Si}{1} ($\lambda$10730.3, 10752.3, 10787.5, 10789.8, 10871.8,
10872.5), 
\ion{Mg}{1} $\lambda$10814.1, and \ion{C}{1} $\lambda$10732.5.
For stars with spectral type of K3 or later, the
\ion{Fe}{1} lines ($\lambda$10756.0, 10786.0)  were
used in addition to the above lines;
however, \ion{C}{1} was not used because it was weak in case of
later-type stars. 
We further derived $v_{\rm broad}$, RVs, and $\gamma_Y$ by comparing
several synthetic spectra. The best fit was determined using the minimum
chi-squared value by varying the three parameters of $v_{\rm broad}$,
RV, and $\gamma_Y$.
The synthetic spectra were constructed using the following limits on
each parameter: (1) $v_{\rm broad}$ values around initially estimated
values with 5\,km\,s$^{-1}$ step; (2) $\gamma_Y$ values in the range of
0.0 to 2.0 with 0.1 step; and (3) observed spectra with wavelength
shifts for RVs with 0.1\,\AA\ step.
For the fitting, the same wavelength range was used as was used in the
initial estimate of $v_{\rm broad}$.
When estimating $v_{\rm broad}$ and RV, we also estimated $\gamma_J$
using spectra with order $m=44$ in the same way as the derivation of
$\gamma_Y$ but in the spectral range of $\lambda = 12620$ to 12700 using
lines of \ion{Fe}{1} $\lambda$12642.2, \ion{Fe}{1} $\lambda$12652.2, and
\ion{Na}{1} $\lambda$12682.7.
The obtained properties ($v_{\rm broad}$, RV, $\gamma_Y$, and
$\gamma_J$) are summarized in Table~\ref{tab:prop_W}.
Finally, we obtained residual profiles of P$\beta$ and \ion{He}{1} by
subtracting synthetic spectra considering the obtained properties from
observed spectra. The profiles relative to the stellar rest velocities
are shown in Figure~\ref{fig:residual}.

We note that we could not estimate $v_{\rm broad}$, RV, and veilings
using the above procedure for three early-type stars with spectral types
of late B--early A: V892 Tau, AB Aur, and IRAS 0401+3103.
This was due to a lack of prominent photospheric features in their
spectra except for \ion{H}{1}.
There are only a small number of photospheric features in the spectra
for such early-type stars, and even the existing features become quite
broadened and very shallow due to the large rotation velocities (e.g.,
$v \sin i = 116$\,km\,s$^{-1}$ for AB Aur; 100\,km\,s$^{-1}$ for V897
Tau).
Instead, we used synthetic spectra assuming $v_{\rm broad}$ and RV
values from the literature, without considering veilings ($\gamma_Y =
0$,$\gamma_J = 0$), for the subtraction of photospheric features from
the observed spectra.
As for $v_{\rm broad}$ of AB Aur, only the projected rotation velocity
($v \sin i$) was provided by the literature \citep{Alecian2013}, $v \sin
i = 100$\,km\,s$^{-1}$.
Nonetheless, we had to input $v_{\rm broad}$, using combinations of $v
\sin i$ and macroturbulence velocity ($v_{\rm mac}$), when making
synthetic spectra with SPTOOL.
\citet{Alecian2013} assumed a $v_{\rm mac}$ of 2\,km\,s$^{-1}$, and
suggested that in the case that $v \sin i$ is larger than
40\,km\,s$^{-1}$, $v_{\rm mac}$ is not a significant
parameter.
Therefore, we adopted their $v \sin i$ values for $v_{\rm broad}$. 
Also, for the $v_{\rm broad}$ value for V892 Tau, only $v \sin i$
(116\,km\,s$^{-1}$) was provided by the literature \citep{Mooley2013}.
Although we could not find the adopted value of $v_{\rm mac}$, it should
 be negligible for $v_{\rm broad}$, considering the large $v \sin i$
 value.
Therefore, we adopted this value as $v_{\rm broad}$.
However, we could not find any literature references for the $v_{\rm
broad}$ and RV values for IRAS 04101+3103.
We adopted the average value of RVs for all target objects in this paper
(Table~\ref{tab:target} Col. (9)) that all exist in the Taurus
star-forming region,  $\sim$20\,km\,s$^{-1}$.
For $v_{\rm broad}$, we assumed a value of 100\,km\,s$^{-1}$, similar to
those for AB Aur and V892 Tau, considering that their spectral types are
almost the same and their disks evolutionary phases are the same (Phase
I).
As for veilings, we could not find any references about veilings for
these three sources. Because the photospheric \ion{He}{1} lines show
prominent features only in very early-type stars, $\sim$B5 and earlier
\citep{Cauley2014}, which are not included in our targets, variations of
synthetic spectra have little impact on our residual profiles.
However, because photospheric P$\beta$ lines show very prominent
absorption features in late B- and early A-type stars (our early-type
targets), differences between photospheric features in assumed synthetic 
spectra and those in actual spectra may have a great impact on the
residual profiles in normalization. We discuss the possible impact of
assuming zero veilings on obtained residual profiles in
Section~\ref{sec:Pbeta}.

Estimated $v_{\rm broad}$, RV, $\gamma_Y$, and $\gamma_J$ values for
each target are summarized in Table~\ref{tab:prop_W}.
WTTSs, comparable to Phase III objects in this paper, are known to show
no veiling \citep[e.g.,][]{Edwards2006}, while correlation between mass
accretion rate and veiling for CTTSs is suggested
\citep[e.g.,][]{Beristain2001}. 
For all four Phase III objects (HP Tau/G2, HD 283572, HBC 388, and V410
Tau), $\gamma_Y$ and $\gamma_J$ are estimated to be almost zero
although some estimated values are non-zero ($\gamma_Y = \gamma_J = 0.1$
for HP Tau/G2; $\gamma_J = 0.1$ for HD 283572).
Therefore, taking all results of Phase III objects into an account,
estimated veilings are $\le$0.1, suggesting that veilings are accurately
estimated with uncertainties of $\sim$0.1.


\subsection{P$\beta$ line profiles} \label{sec:Pbeta}

Paschen line emissions are thought to arise primarily in the
magnetospheric accretion column \citep{Hartmann2016}.
Although profiles of Paschen lines in obtained spectra (P$\epsilon$,
P$\gamma$, P$\delta$, and P$\beta$) show similar morphologies, we
discuss those of P$\beta$ that show the most prominent features, as
discussed in \S~\ref{sec:Paschen}.
In this section, we focus on the properties of P$\beta$ emission for
each source.
We discuss the implications for the profile morphology in
Section~\ref{sec:discussion}.

\begin{figure*}[h]
\includegraphics[scale=0.7]{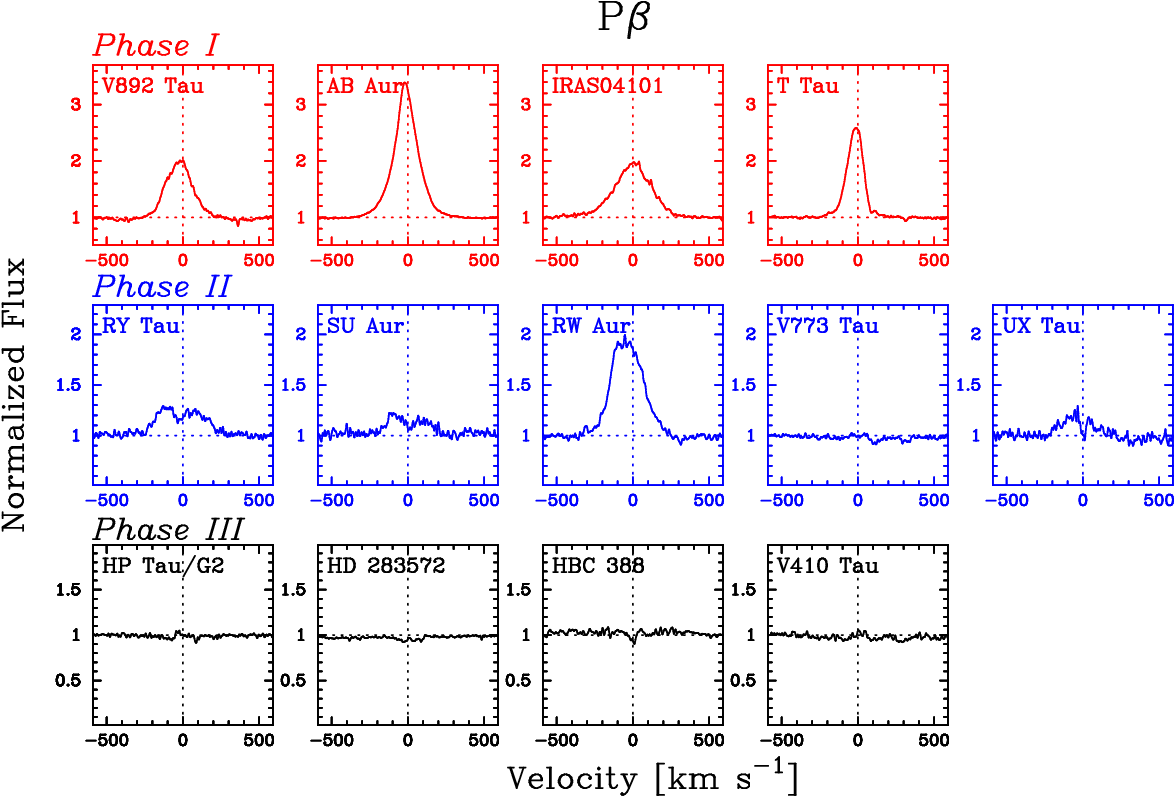}

 \vspace{3em}
 \clearpage
 \includegraphics[scale=0.7]{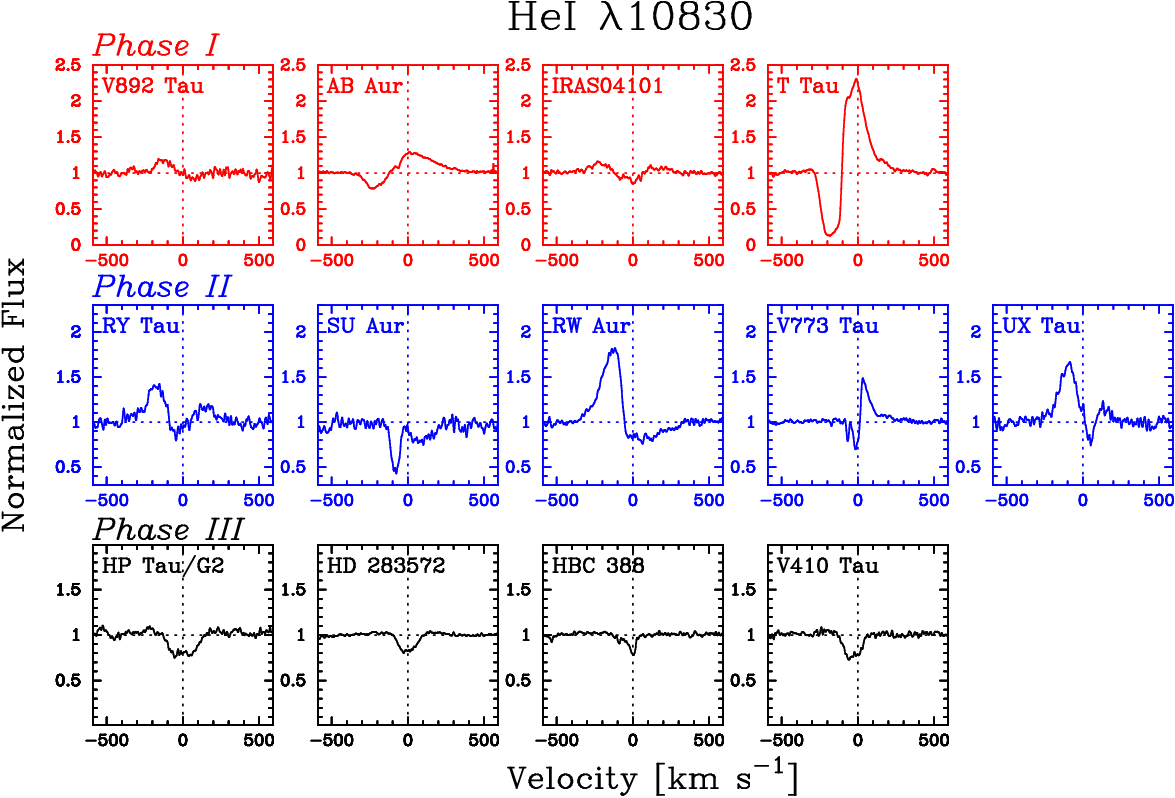}
 
\caption{Residual P$\beta$ profiles (top) and \ion{He}{1} profiles
(bottom).
Velocities are relative to the stellar rest velocities. Spectra for
 Phase I, II, and III sources are shown in top, middle and bottom
 panels, respectively.}  \label{fig:residual}
\end{figure*}

We detected P$\beta$ in 8 of our 13 targets (all Phase I sources, and
all Phase II sources except V773 Tau), while these lines are not
detected in 5 sources (V773 Tau and all Phase III sources).
The detection rates for Phase I, II, and III objects are 100\,\% (4/4),
80\,\% (4/5), and 0\,\% (0/4), respectively.
In Table~\ref{tab:profile_Pbeta}, we summarize the P$\beta$ profile
parameters obtained from the residual profiles
(Figure~\ref{fig:residual}), showing object name in column 1, profile
type in column 2, the maximum blueshifted and redshifted line velocities
($V^{\rm blue}_{\rm max}$ and $V^{\rm red}_{\rm max}$) in columns 3 and
4, the velocity of peak emission and absorption ($V^{\rm em}_{\rm peak}$
and $V^{\rm abs}_{\rm peak}$) in columns 5 and 6, the line fluxes
relative to the continuum at the peak emission and absorption velocities
($F^{\rm em}_{\rm peak}$ and $F^{\rm abs}_{\rm peak}$) in columns 7 and
8, emission and absorption equivalent widths (Em. $W_\lambda$ and
Ab. $W_\lambda$) in columns 9 and 10, and the type of subcontinuum
absorption in column 11.
In column 2, we categorize residual spectra into morphology groups based
on the classification scheme of
\citet{Cauley2014}.
They divided the line profiles into six groups: (1) P-Cygni (PC), (2)
inverse P-Cygni (IPC), (3) pure absorption (A), (4) single-peaked
emission (E), (5) double-peaked emission (DP), and (6) featureless (F).
We added a new category, a profile with both blueshifted and redshifted
subcontinuum absorption but with zero or one emission peak (hereafter,
``BR''). 
\citet{Reipurth1996} has reported another well-known classification.
The groups E, PC, and IPC in the classification by \citet{Cauley2014}
correspond to type I, type IV B, and type IV R in the classification of
\citet{Reipurth1996}, respectively, and that group DP by
\citet{Cauley2014} comprises type II and type III of
\citet{Reipurth1996}.

The P$\beta$ profiles for the targets are categorized into the following
four groups;
group E, group DP, group IPC, and group F.
The profiles of all the four Phase I sources
(V892 Tau, AB Aur, \IRAS, and T Tau) are categorized into group E.
They all exhibit approximately symmetric and broad features with
$|V^{\rm em}_{\rm peak}| \le 20$\,km\,s$^{-1}$ for all the sources, and
$|V^{\rm blue}_{\rm max}|$ and $|V^{\rm red}_{\rm max}|$ of
$\gtrsim$300\,km\,s$^{-1}$ except for T Tau ($|V_{\rm max}| \sim
250$\,km\,s$^{-1}$).
They have slightly redward absorption features judging from $|V^{\rm
blue}_{\rm max}|$ values exceeding $|V^{\rm red}_{\rm max}|$ by
$\sim$50\,km\,s$^{-1}$.
In addition, the profiles of three of the sources 
(RY Tau, SU Aur, and UX Tau) are categorized into group DP.
All the sources in this group have influence from redward absorption, 
judging from $V^{\rm em}_{\rm peak} < 0$. 
The profile for UX Tau shows subcontinuum absorption
centered on the stellar rest velocity.
The profile of RW Aur is categorized into group IPC, which
has a very slight subcontinuum feature with $F^{\rm abs}_{\rm peak} =
0.96$.
Finally, the profiles of the remaining five sources (V773 Tau, HP Tau/G2, HD 
283572, HBC 388, and V410 Tau) are categorized into group F.

We note the possibility that photospheric features in P$\beta$
profiles for A- and B-type stars (V892 Tau, AB Aur, and \IRAS) are not
completely removed in residual profiles shown in Fig.~\ref{fig:residual}
because there is no information on veilings for the early type stars
and we tentatively adopted synthetic spectra assuming zero veilings as
photospheric features (Section~\ref{sec:veil_PsF}).
We checked the variation of obtained residual profiles assuming the
extreme case of high veilings, 2.0, as shown in
Table~\ref{tab:profile_Pbeta} in parentheses.
As a result, we found that variations are $\sim$100\,km\,s$^{-1}$.
Although the variations are very large, their impacts are negligible
because the interpretation of the profile is only based on line
morphology and not line width or intensity
(Section~\ref{sec:interpret_line}).

\subsection{\ion{He}{1} $\lambda$10830 line profiles} \label{sec:HeI} 

\ion{He}{1} $\lambda$10830 line emissions are thought to have a
composite origin, including contributions from wind, from the funnel
flow, and from an accretion shock \citep{Edwards2006}.  In this section,
we focus on the properties of \ion{He}{1} emission for each source. 
We discuss the implication of the profile morphology in
Section~\ref{sec:discussion}.

We detect \HeI $\lambda$10830 features with absorption below the
continuum in all of our targets.
The detection rates of \ion{He}{1} lines for Phase I, II, and III
objects are 100\,\% (4/4), 100\,\% (5/5), and 100\,\% (4/4),
respectively. 
In Table~\ref{tab:profile_HeI}, we summarize the \ion{He}{1}
profile parameters that are obtained from residual profiles in
Fig.~\ref{fig:residual}, in the same manner as for P$\beta$ profile
parameters in Section~\ref{sec:Pbeta}.
In column 2, we categorize \ion{He}{1} residual spectra into morphology
groups in Section~\ref{sec:Pbeta}.

The \ion{He}{1} profiles for the targets are categorized
into five groups, group IPC, group PC, group DP, group BR, and group A.
The profiles of two sources (V892 Tau and RW Aur) are categorized into
group IPC; both exhibit broad features with $|V^{\rm blue}_{\rm max}|$
and $|V^{\rm red}_{\rm max}|$ of $\gtrsim$250\,km\,s$^{-1}$.
The profiles of three of the sources (AB Aur, T Tau, and V773 Tau) 
are categorized into group PC; 
the profiles of AB Aur and T Tau exhibit broad features with
$|V^{\rm blue}_{\rm max}|$ and 
$|V^{\rm red}_{\rm max}| \gtrsim 250$\,km\,s$^{-1}$, while that of 
V773 Tau has relatively narrow features with $|V^{\rm blue}_{\rm max}|$ and
$|V^{\rm red}_{\rm max}| \lesssim 250$\,km\,s$^{-1}$.
The profiles of three of the sources (IRAS 04101+3101, RY Tau, and UX
Aur) are categorized into group DP; they exhibit broad features with
$|V^{\rm blue}_{\rm max}|$ and $|V^{\rm red}_{\rm max}|$ of
$\gtrsim$250\,km\,s$^{-1}$.
The profiles of all the four Phase III sources (HP Tau/G2, HD 283572,
HBC 388, and V410 Tau) are considered to be group A; they demonstrate
centered subcontinuum and exhibit narrow profiles with $|V^{\rm
blue}_{\rm max}|$ and $|V^{\rm red}_{\rm max}|$ of
$\lesssim$150\,km\,s$^{-1}$.
The profile of SU Aur is assigned to a new category, group BR; it
exhibits redshifted and blueshifted subcontinuum absorption, which is
relatively narrow feature with $|V^{\rm blue}_{\rm max}|$ and $|V^{\rm
red}_{\rm max}| \lesssim 250$\,km\,s$^{-1}$.

\section{DISCUSSION} \label{sec:discussion}

Our WINERED data present a variety of profile morphologies in P$\beta$
and \ion{He}{1} $\lambda$10830 lines.
In this section, 13 sources observed in this study and the six sources
observed in previous studies ({among them five sources each for P$\beta$
and \ion{He}{1}}; \citealt{Folha2001}, \citealt{Edwards2006}; see
Section~\ref{sec:targets}) are considered in combination for obtaining
improved statistical accuracy. 
The statistics of the categorized line profile morphologies
(Section~\ref{sec:profile}) and the manner in which the profile
morphology for each target provides an insight into magnetospheric
accretion and inner winds (Section~\ref{sec:interpret_line}),
and the possible progression of dominant processes thorough the disk
evolutionary phases (Section~\ref{sec:Progression}) are discussed
herein.
The suggested progression is compared with previous studies
(Section~\ref{sec:comparison}).
The chromospheric activities in Phase III sources from \ion{He}{1}
profile morphologies {are also evaluated} (Section~\ref{sec:chrom_III}).
Finally, the implications of the results are considered for the theories
of mass flow processes (Section~\ref{sec:Implication}).

\subsection{Profile morphologies} \label{sec:profile} 

\subsubsection{P$\beta$ line profile morphologies} \label{sec:profile_Pb}

Among the six additional stars, \citet{Folha2001} presented the P$\beta$
profiles for five of them (CW Tau, HP Tau, GM Aur, DR Tau, and DS Tau). 
The P$\beta$ line profile morphologies for 18 intermediate-mass stars
(the 13 targets observed in the present study and the five additional
targets observed in previous studies) are considered in combination.
Based on the classification scheme in Section~\ref{sec:Pbeta}, they are
categorized into two groups: two (CW Tau and DR Tau) in group DP and
three (HP Tau, GM Aur, and DS Tau) in group IPC (see
Table~\ref{tab:target_others}).

The statistics of the P$\beta$ line profile morphologies for 18
intermediate-mass stars for each evolutionary phase are summarized in
Table~\ref{tab:stat_Pbeta}.
The P$\beta$ line profiles for Phase I sources are categorized into three
groups: four (V892 Tau, AB Aur, IRAS 04101+3103, and T
Tau) in group E;
two (CW Tau and DR Tau) in group DP; and 
one (HP Tau) in group IPC. 
The P$\beta$ line profiles for Phase II sources are categorized into
three groups: three (RY Tau, SU Aur, and UX Tau) in group DP;
three (RW Aur, GM Aur, and DS Tau) in group IPC; and 
one (V773 Tau) in group F. 
Finally, the P$\beta$ line profiles of all of the Phase III sources are
categorized into group F.
The statistics of Phase III are clearly different from those of Phase I
and II, while those of Phase I and Phase II do not appear to differ
significantly, except for the fact that a large fraction of Phase I
sources but none of the Phase II sources are present in group E.
\citet{Edwards2006} reported the P$\gamma$ profiles for five sources (CW Tau,
GM Aur, DR Tau, DS Tau, and HN Tau) among the six additional sources 
noted in Section~\ref{sec:targets}. 
The trends observed in this profiles are similar to those in the 
P$\beta$ profiles although the P$\gamma$ profile features are sometimes 
weaker than the P$\beta$ features (Section~\ref{sec:Paschen}):
of the Phase I sources, among which two are in group E and one is 
in group IPC, and of the Phase II sources, among which one is in group
F and one is in group IPC. 
However, it should be noted that the profile types are not necessarily
the same for each source between P$\beta$ profiles presented by
\citet{Folha2001} and P$\gamma$ profiles presented by
\citealt{Edwards2006}.
Considering that the spectra of P$\gamma$ and P$\beta$ show similar
features across various targets in the same observation period (see
Section~\ref{sec:Paschen}), the differences of the profile types can be
attributed to the differences in the observation period;
this will be further discussed in Section~\ref{sec:Progression}.

We compared P$\beta$ profile morphologies for intermediate-mass stars
and those from previous Paschen line spectroscopic observations of YSOs
with high spectral resolution \citep{{Folha2001}, {Edwards2006}}.
These studies' targets were primarily low-mass\footnote{Some
of the targets by \citet{Folha2001} and \citet{Edwards2006} are the same
as ours: V773 Tau, RY Tau, T Tau, SU Aur, and RW Aur in
\citet{Folha2001}; SU Aur, RW Aur{, and UX Tau} in
\citet{Edwards2006}.}  T Tauri stars.
\citet{Folha2001} presented P$\beta$ line profiles for 49 T Tauri stars
(44 CTTSs, four WTTSs, and one variable star between CTTS and WTTS).
Among them, 45 sources (41 CTTSs, three WTTSs, and one variable star
between CTTS and WTTS) with sufficient S/Ns are counted later in this
paper\footnote{\citet{Folha2001} pointed out that for some
sources P$\beta$ lines are too noisy to allow a reliable classification
of their profiles: three CTTSs (DQ Tau, FX Tau and GH Tau), and one WTTS
(DI Tau).}.
\citet{Edwards2006} presented P$\gamma$ line profiles for 48 T Tauri
 stars (39 CTTSs, 6 WTTSs, and three Class I sources).
Although the observations by \citet{Edwards2006} were not for P$\beta$,
the P$\gamma$ morphology of the profile is known to be reminiscent of
the P$\beta$ profile \citep{Edwards2006}.
\citet{Folha2001} detected P$\beta$ emission in 38 of 41 CTTSs in their
sample, while \citet{Edwards2006} detected P$\gamma$ emission in 38 of
39 CTTSs in their sample.
For profiles of low-mass stars from previous studies, redshifted
absorption features are common, while none shows blueshifted absorption
features. This is the case with obtained profiles for intermediate-mass 
stars in our sample.
For comparison, we show the statistics of profile morphology of Paschen
lines for CTTSs by \citet{Folha2001} and \citet{Edwards2006} in
Table~\ref{tab:stat_Pbeta}.
For both samples, the occurrence rate of group E is highest ($\sim$50 to
70\,\%). 
This tendency is similar to that for Phase I sources in 
the intermediate-mass star sample.
Considering that CTTSs are comparable to Phase I
(Section~\ref{sec:targets}), this tendency seems to be a natural
consequence.
The occurrence rate of group IPC sources is the second
highest for CTTSs in previous studies conducted by
\citet{Folha2001} and \citet{Edwards2006}, $\sim$20 to 30\,\%.
This is not inconsistent with those for Phase I and II sources in the
intermediate-mass star sample.
Meanwhile, group DP sources are very rare in CTTSs ($\lesssim$10\,\%),
as suggested by \citet{Folha2001}, while occurrence rate is relatively
high for Phase I and II sources in the intermediate-mass star sample
($\sim$30--40\,\%).
This may be characteristics for intermediate-mass stars.
However, the statistics of CTTSs do not appear to differ significantly
from those of Phase I and II sources in the intermediate-mass star
sample.
Lastly, we also show the statistics of profile morphology of Paschen 
lines for WTTSs, which are comparable to low-mass stars in Phase III, by
\citet{Folha2001} (P$\beta$) and \citet{Edwards2006} (P$\gamma$) in
Table~\ref{tab:stat_Pbeta}.
Paschen lines are undetected in almost all WTTSs, both in
\citet{Folha2001} and in \citet{Edwards2006}, prompting a classification
in group F.
This result is consistent with those for our target intermediate-mass
stars in Phase III.

\subsubsection{\ion{He}{1} line profile morphologies} \label{sec:profile_Pb}
\label{sec:profile_HeI}

Among the six additional stars, \citet{Edwards2006} reported the
\ion{He}{1} profiles for five of them (CW Tau, GM Aur, DR Tau, DS Tau,
and HN Tau). 
In this section, the \ion{He}{1} line profile morphologies for 18
intermediate-mass stars (the 13 targets observed in the present study
and the five additional targets observed in previous studies are
considered.
The sources are categorized into four groups based on the classification
scheme in Section~\ref{sec:Pbeta}: 
two (CW Tau and HN Tau) are assigned to group DP,
one (GM Aur) is assigned to group F,
one (DR Tau) is assigned to group PC,
and one (DS Tau) is assigned to group BR
(see Table~\ref{tab:target_others}).

The statistics of the \ion{He}{1} line profile morphologies for the 18
intermediate-mass stars for each evolutionary phase are shown in
Table~\ref{tab:stat_HeI}.
The line profiles of Phase I sources are categorized into three groups:
three (AB Aur, T Tau, and DR Tau) are categorized into group PC, 
three  (IRAS 04101+3103, CW Tau, and HN Tau) are in group DP, 
and one (V892 Tau) is in group IPC. 
The Phase II sources are categorized into five groups: 
two (RY Tau and UX Tau) in group DP,
two (SU Aur and DS Tau) in group BR, 
one (RW Aur) in group IPC, 
one (V773 Tau) in group PC,
and one (GM Aur) in group F. 
Finally, all the Phase III sources are categorized into group A.
Based on these statistics, the \ion{He}{1} profiles of the sources in
Phase I and II are similar to each other but the profiles of the Phase
III sources are clearly different (all in group A) from those in the
remaining phases.
However, in terms of broadness of spectral features, there seem to be
clear differences between Phase I and II sources:
all four Phase I sources show broad features with $|V^{\rm blue}_{\rm
max}|$ or $|V^{\rm red}_{\rm max}|$ of $\gtrsim$250\,km\,s$^{-1}$,
while some Phase II sources (SU Aur and V773 Tau) show narrow features
with $|V^{\rm blue}_{\rm max}|$ or $|V^{\rm red}_{\rm max}|$ of
$\lesssim$250\,km\,s$^{-1}$.

We compare \ion{He}{1} $\lambda$10830 profile morphologies for
intermediate-mass stars 
with those from previous
\ion{He}{1} spectroscopic observations of YSOs with high spectral
resolution. 
For these comparisons, we reference \citet{Oudmaijer2011} and
\citet{Cauley2014} for HAeBes (corresponding to intermediate-mass stars
in Phase I) and \citet{Edwards2006} mainly for low-mass T Tauri stars.
As for intermediate-mass stars, \citet{Oudmaijer2011} detected
\ion{He}{1} features in 79 out of 90 of their sample HAeBes. 
Although all of their spectra are not shown in their paper, the
partially shown observed spectra (not residual profiles) show group PC,
IPC, or DP profiles.
\citet{Cauley2014} observed 56 HAeBes, 28 HAe stars and 28 HBe
stars. For HAe stars, whose mass range is roughly comparable to that of
our targets in this paper, they detected \ion{He}{1} features in 26
stars.
We summarize statistics of profile morphology for their targets in
Table~\ref{tab:stat_HeI} based on their Fig.~4.
As a result, the occurrence frequencies of groups PC and IPC are
relatively higher compared to those of the other types. 
The statistics of HAe stars are not inconsistent with those of Phase I
and Phase II sources in the intermediate-mass star sample in this study. 
\citet{Cauley2014} pointed out that
narrow blueshifted absorption features are not seen in HAe stars in
their sample while broad blueshifted absorption features are common.
This tendency is also seen in our Phase I sources. Considering that
narrow blueshifted absorption features are seen only in some Phase II
objects in the sample of intermediate-mass stars (RY Tau, SU Aur, V773
Tau, and DS Tau), these features may be characteristics for Phase II
objects.

As for low-mass stars, \citet{Edwards2006} detected \ion{He}{1} features
in 38 of 39 CTTSs in their sample.
We show the statistics of \ion{He}{1} profile morphology for low-mass
stars from \citet{Edwards2006} in Table~\ref{tab:stat_HeI}.
CTTSs are categorized into almost all profile types.
The occurrence frequency of group PC is highest, while that of group BR
is second highest.
Group PC sources are observed among both Phase I and II sources in the
intermediate-mass star sample, but group BR sources are only observed
among Phase II sources.
However, the statistics of CTTSs do not appear to differ significantly
from those of Phase I and II sources in the intermediate-mass star
sample. 
\citet{Edwards2006} suggested that broad P Cygni-like profiles are more
common among stars of high veilings ($\gamma_Y \ge 0.5$), while narrow 
emission coupled with both blueshifted and redshifted absorption is more
common among stars with low veilings ($\gamma_Y \le 0.2$).
This tendency aligns with the results of the intermediate-mass samples
for \ion{He}{1} profiles:
group PC sources are more common in Phase I, while some sources with
narrow emission and redshifted and blueshifted absorption features are
seen only in Phase II.
On the contrary, we also show the statistics of \ion{He}{1} profile
morphology for WTTSs by \citet{Edwards2006}, which are comparable to
low-mass stars in Phase III, in Table~\ref{tab:stat_HeI}.
The highest occurrence frequency belongs to group F (featureless) for
four sources, and the second highest is group E (pure emission) for two 
sources.
This result is not consistent with those for our target
intermediate-mass stars in Phase III, which will be discussed later
(Section~\ref{sec:PhaseIII}).

\subsection{Interpretation of P$\beta$ and \ion{He}{1} line profile
  morphologies} \label{sec:interpret_line}

Paschen lines exhibit a centered emission peak and a redshifted
absorption that can be used as diagnostics for magnetospheric accretion, 
while \HeI profiles show both blueshifted and redshifted absorptions
that can be used as diagnostics for magnetospheric accretion flows and
inner winds, respectively \citep{Hartmann2016}.
We largely based our interpretations on profile morphology of P$\beta$
and \ion{He}{1} line profile rather than line width or intensity.

\subsubsection{Magnetospheric accretion} \label{sec:MA}

Both P$\beta$ and \HeI line profiles show redshifted absorptions that
can be used as diagnostics for magnetospheric accretion
(e.g., \citealt{Folha2001}, \citealt{Edwards2006}, \citealt{Fischer2008},
\citealt{Kurosawa2011}).
First, we diagnose the sources in which sources magnetospheric
accretions are functioning using the P$\beta$ line profiles.
The line profiles for RW Aur, HP Tau, GM Aur, and DS Tau show redshifted
subcontinuum features, displaying a clear magnetospheric accretion
signature.
For nine sources (V892 Tau, AB Aur, IRAS 04101+3103, T Tau, RY Tau, SU
Aur, UX Tau, CW Tau, and DR Tau), the observed line profiles show
emission features but do not show redshifted subcontinuum features.
Among them, four sources (V892 Tau, AB Aur, IRAS 04101+3103, and T Tau)
are categorized into group E (pure emission), whereas five sources (RY
Tau, SU Aur, UX Tau, CW Tau, and DR Tau) are categorized into group DP
(double-peaked emission).
Considering the sources in group E have $|V^{\rm blue}_{\rm max}| >
|V^{\rm red}_{\rm max}|$ and that the sources in group
DP{, except for CW Tau,} have $V^{\rm em}_{\rm peak} <
0$, they show slight influence from redward absorption features,
suggesting that magnetospheric accretion activities are functioning.
The model Br$\gamma$ line profiles\footnote{Br$\gamma$ line profiles are
known to show very similar features to P$\beta$ line profiles
\citep{Folha2001}.}  for sources with magnetospheric accretion
calculated by \citet{Muzerolle1998a} also show a redward absorption
feature (not subcontinuum features) at inclination angles of
$\lesssim$60$^\circ$ (see their Fig.~5).
Therefore, the P$\beta$ observed line profiles suggest that
magnetospheric accretion processes are functioning in the
eight 
sources {that exhibit
emission features but not redshifted subcontinuum features, except for
CW Tau,} in addition to four sources that do exhibit
redshifted subcontinuum features.

\ion{He}{1} can also be used as diagnostics for magnetospheric
accretion.
The observed line profiles for five sources (V892 Tau, SU Aur, RW Aur,
UX Tau, and DS Tau) show redshifted subcontinuum features, displaying a
clear magnetospheric accretion signature.
For the other sources in Phase I and II{, except for GM 
Aur}, the line profiles show \ion{He}{1} emissions but do not show the
redshifted subcontinuum features. This seems to suggest that the
activities of magnetospheric accretion are not functioning.
However, \citet{Kurosawa2011} indicated that wind-related \ion{He}{1}
emissions are relatively minor, whereas the magnetosphere is a main 
emission contributor (see Fig.~12 in \citealt{Kurosawa2011}).
The line profiles for three sources (RY Tau, CW Tau, and HN Tau),
which are categorized into group DP, also show redward absorption
features, judging from $V^{\rm em}_{\rm peak} < 0$, suggesting
magnetospheric accretion activities.
Therefore, \ion{He}{1} profiles for all Phase I and II sources, except
for GM Aur, suggest magnetospheric activities.

In summary, all Phase I and II sources suggest magnetospheric activities
from the observed line profiles of P$\beta$ and \ion{He}{1}
although the profile for approximately half of them
(V892 Tau, SU Aur, RW Aur, UX Tau, HP Tau, GM Aur, and DS Tau) show
redshifted subcontinuum features in P$\beta$ and/or \ion{He}{1} lines,
which are clear magnetospheric accretion signatures.

\subsubsection{Inner winds} \label{sec:IW}

\HeI profiles show both blueshifted and redshifted absorptions that can 
be used as diagnostics for magnetospheric accretion flows and inner
winds, respectively \citep{Hartmann2016}.
\citet{Kwan2007} modeled blueshifted absorption at \ion{He}{1}
$\lambda$10830 in CTTS via scattering of the stellar and veiling
continua as a probe of inner winds,
while \citet{Kurosawa2011} presented multidimensional non-local
thermodynamic equilibrium radiative transfer models of hydrogen and
helium line profiles formed in the magnetospherical accretion and inner
winds of CTTSs, including \HeI $\lambda$10830 and P$\beta$. 
Both models are consistent with the scenario in which the narrow
blueshifted absorption component of \HeI $\lambda$10830 seen in
observations is caused by a disk wind, and the wider blueshifted
absorption component (the terminal velocity of the wind up to
400\,km\,s$^{-1}$) is caused by stellar wind.

In the sample of intermediate-mass stars, the line profiles for seven
sources (AB Aur, RY Tau, SU Aur, T Tau, V773 Tau, DR Tau, and DS Tau)
have blueshifted subcontinuum absorption features.
Among them, the line profiles for AB Aur, T Tau, and DR Tau are
categorized into group PC with broad blueshifted subcontinuum absorption
($V^{\rm blue}_{\rm max} \sim -$300--$-$400\,km\,s$^{-1}$).
They are consistent with the stellar wind case. 
The line profiles for RY Tau, SU Aur, V773 Tau, and DS Tau show narrow
blueshifted subcontinuum absorption features, with terminal wind
volocities of $\lesssim$150\,km\,s$^{-1}$.
They are consistent with the disk wind case.
The stellar wind is suggested to be functioning in AB Aur, T Tau, and DR
Tau, whereas the disk wind is suggested to be functioning in RY Tau, SU
Aur, V773 Tau, and DS Tau.

\subsubsection{Phase III sources}
\label{sec:PhaseIII}

The observed P$\beta$ profiles show group F features (featureless),
while the observed \HeI profiles for all Phase III sources show group A
features (pure absorption) with centered subcontinuum absorption.
These features are not expected in model profiles in the case of disk
wind, or stellar wind (e.g., \citealt{Kurosawa2011}). 
This suggests that these sources have no such activities.
Moreover, because absorption features in \HeI are broader for sources with
larger $v_{\rm broad}$, the lines are more likely to be stellar
intrinsic features than circumstellar features.


\subsection{Progression of dominant processes with evolution of 
protoplanetary disks} \label{sec:Progression}

In Section~\ref{sec:interpret_line}, observed P$\beta$ and \ion{He}{1}
profiles for each source were interpreted. 
In Phase I, three sources (AB Aur, T Tau, and DR Tau) show features of
stellar wind, while the other four sources (V892 Tau, IRAS 04101+3103{,
CW Tau, and HN Tau}) show no features of inner winds.
For magnetospheric accretion, only V892 Tau and HP Tau show clear
features, while the other sources show features suggesting the activity.
In Phase II, four sources (RY Tau, SU Aur, V773 Tau, and DS Tau) show
features of disk wind, while the other three sources (RW Aur, UX Tau,
and GM Aur) show no features of inner winds.
For magnetospheric accretion, five sources (SU Aur, RW Aur, UX Tau{, GM
Aur, and DS Tau}) show clear features, while the other two sources show
the feature suggesting the activities.
In Phase III, no targets show activities of mass accretion, disk winds,
or stellar winds.
This suggests that dominant processes in different evolutionary phases
are different, particularly for inner winds. This finding indicates a
progression of dominant mass flow processes with disk evolution:
stellar wind and probably magnetospheric accretion in Phase I,
magnetospheric accretion and disk wind in Phase II, and no activities in
Phase III. 
In particular, progression is observed for
inner winds. The statistics of the blueshifted absorption types for
\ion{He}{1} lines, broad or narrow, which are used for diagnostics of
dominant inner winds, are summarized in Table~\ref{tab:stat_HeI}.

\begin{figure*}[h]
\includegraphics[scale=0.5]{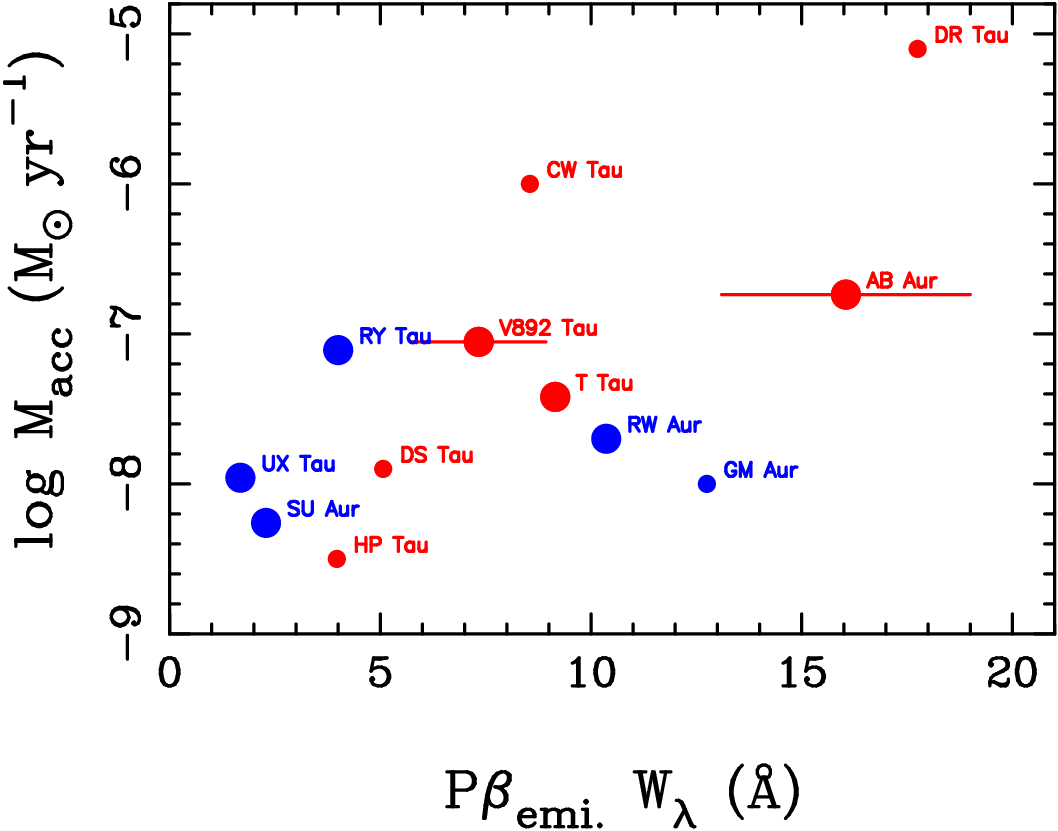}
   \hspace{1em}
\includegraphics[scale=0.5]{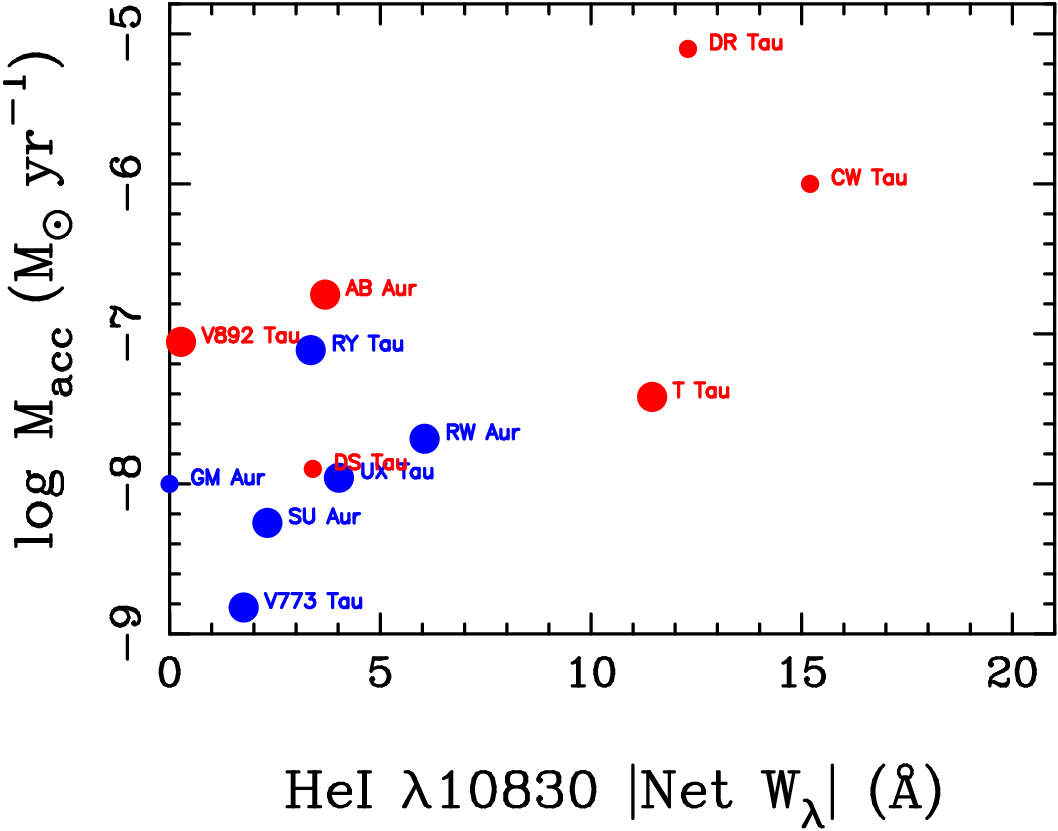}
\caption{Equivalent width vs. mass accretion rate.
The target intermediate-mass stars in this study are
presented by large filled circles. Phase I, II, and III sources are
shown with red, blue, and black, respectively.
Six additional intermediate-mass stars in the samples of \citet{Folha2001} 
and \citet{Edwards2006} are denoted by small symbols with the same colors as 
that of the targets observed in the present study. 
The mass accretion rates for the targets observed in the present study
are shown in Table~\ref{tab:disk_prop}, that for HP Tau is reported by
\citet{Johns-Krull2002}, and those for the five remaining additional
sources are reported by \citet{Edwards2006}.
Left: P$\beta$ emission equivalent width and mass accretion rate.
Equivalent widths for the targets observed in the present study are
shown in Table~\ref{tab:profile_Pbeta}. 
The error bars of Phase I sources are for V892 Tau and AB Aur, showing
the equivalent widths in the case of $\gamma = 0.0$ and 2.0
(Section~\ref{sec:Pbeta}).
The equivalent widths for the additional sources are from
\citet{Folha2001}; those for HP Tau, GM Aur, and DS Tau (showing PC
profiles) are obtained from Table~6 in that article and those for CW Tau
and DR Tau are roughly estimated from Figure~1 in that article. 
Right: Sum of absolute values of emission plus absorption \ion{He}{1}
equivalent widths vs. mass accretion rate.
The equivalent widths for the targets observed in the present study are
shown in Table~\ref{tab:profile_HeI} and those for the additional
sources are from Table~1 of the study by \citet{Edwards2006}.}
\label{fig:EW}
\end{figure*}

However, some points still need to be considered. 
i) YSOs tend to be variable, which can make line
 profiles differ among observation periods \citep{Bouvier2003}.
Previous studies suggest that the general features of {the} observed
line profiles do not change between periods although the lines can show
$\sim$10\,\% variabilities (e.g., \citealt{Kurosawa2005},
\citealt{Edwards2006}). 
However, the times between observations were not very long
(a year at longest in the study conducted by \citet{Edwards2006}). 
On the contrary, the P$\beta$ line profile morphologies reported by
\citet{Folha2001} and the P$\gamma$ line profile morphologies reported
by \citet{Edwards2006} for the same sources show clear difference (e.g.,
those of CW Tau and DR Tau; see Table~\ref{tab:target_others}).
Considering that the P$\gamma$ morphology of the profile is known to be
reminiscent of the P$\beta$ profile (\citealt{Edwards2006}, see also
Section~\ref{sec:Paschen}),
the line profiles can vary by approximately seven rears, reported for
the Paschen series, which has a long time separation (1994--1995 by
\citealt{Folha2001} and 2001--2002 by \citealt{Edwards2006}).
Therefore, continuous observation of \ion{He}{1} is necessary in the
future. 
ii) The targets in this study are limited to only 19 sources.
11 sources are excluded due to limited instrumental sensitivities but
they satisfy other conditions for selecting the target intermediate-mass
stars (Section~\ref{sec:targets}).
A large fraction of the excluded sources are of the latest spectral
type, K5, according to the criteria for selecting the intermediate-mass
stars.
Because each spectral type corresponds to a certain range of star masses
and ages, the sources should have marginal masses between low mass and
intermediate mass \citep{Yasui2014}. Thus, the exclusion may not be
necessarily bad to avoid contaminating the sample of low-mass stars. 
However, because the excluded sources may alter the statistics, more
observations are necessary in the future.
To investigate the universality of the suggested progression, YSOs of a
wide variety of ages and masses should be observed. 
iii) 
The spectral types of Phase I targets
appear earlier (three A- or B-
type and {five} K- type stars) 
than those of Phase II and III targets (G--K
type stars). This bias may impact observed profiles, influencing our
interpretation of dominant activities in each source and phase.
However, such tendencies are not necessarily true for all
intermediate-mass sources observed in this study, e.g., the later
spectral type of T Tau (K0) than those of HP Tau/G2 (G0), RY Tau (G1),
SU Aur (G1), and HD 283572 (G5).
Also, such tendencies are not seen in the six intermediate-mass stars in
the samples of \citet{Folha2001} and \citet{Edwards2006} (see
Section~\ref{sec:targets}).
Therefore, the differences in spectral type should not generate a
systematic bias.
iv) From Tables~\ref{tab:disk_prop} and \ref{tab:prop_W}, Phase I
sources appear to have larger mass accretion rates.
We plotted P$\beta$ and \ion{He}{1} equivalent widths vs. mass accretion 
rate for our intermediate-mass targets with measured mass accretion rate
in Fig.~\ref{fig:EW}.
Sources of Phase I and II are indicated in red and blue, respectively.
P$\beta$ emission equivalent width are shown in the left panel.
The figure suggests that P$\beta$ emission equivalent widths are
moderately correlated with mass accretion rate ($p=0.66$), as suggested
in \citet{Folha2001}, and that Phase I sources seem to have larger mass
accretion.
This may suggest that mass accretion rate determines dominant activities
of inner winds.
However, such tendencies are not necessarily true for all sources, e.g.,
{T Tau (Phase I) has a} smaller mass accretion rate {than that} of
RY Tau (Phase II) {for the targets observed in the present study, and
the mass accretion rates of HP Tau and DS Tau (Phase I) are smaller or
comparable to those for Phase II sources.}
This is also discussed in Section~\ref{sec:comparison} in the context of
previous studies (e.g., \citealt{Edwards2006}).
For reference, we also plotted the sum of absolute values of emission
and absorption of \ion{He}{1} equivalent widths{, which is introduced by
\citet{Edwards2006}, vs. mass accretion rate} (Fig.~\ref{fig:EW}, right
panel).
This shows that the equivalent values of \ion{He}{1} are also correlated
with mass accretion rate ($p=0.75$), which \citet{Edwards2006} specified
using their sample CTTSs.
v) Finally, the weakening of stellar wind signatures and strengthening
 of disk wind signatures in Phase II may be due solely to an overall
 weakening in stellar wind activities through disk evolution.
\citet{Kwan2007} indicated that excitation conditions for \ion{He}{1} 
$\lambda$10830 are more favorable in stellar wind than in disk wind.
In any case, the suggested progression does not change significantly:
stellar winds (and possibly disk winds as well) are active at first, and
then disk winds become relatively active.
However, further quantitative approaches using the fitting of observed
profiles with model profiles are necessary in future work.


\subsection{Comparison of suggested progression of dominant 
processes with previous studies} \label{sec:comparison}

Although we could not find previous results for high-resolution
spectroscopy of young intermediate-mass stars with both \ion{He}{1} and 
P$\beta$, the \ion{He}{1} profiles for HAeBes are presented in
\citet{Cauley2014}.
For HAe stars, whose mass range is roughly comparable to that of our
targets,
the statistics of their P$\beta$ and \ion{He}{1} line profile
morphologies are not significantly different from those of Phase I and
II sources,
as discussed in Section~\ref{sec:profile}. 
\citet{Cauley2014} pointed out that in HAe stars, disk winds (evidenced
by narrow blueshifted absorption features) are not functioning, while
stellar winds (evidenced by commonly broad blueshifted absorption
features) are functioning.
A summary of the statistics of the \ion{He}{1} blueshifted subcontinuum
absorption features for the HAe stars is presented in
Table~\ref{tab:stat_HeI}.
These results are consistent  with our results for Phase I sources. 
Admittedly, our data set has a small sample size (seven Phase I
sources).
But the agreement between the results of their large sample (28 HAe
stars) and those of our smaller sample reinforces the suggestion that
dominant processes for Phase I are stellar wind and probably
magnetospheric accretion.

Previous authors have posited alternative mechanisms for the progression
of dominant processes.
\citet{Edwards2006} mainly focused on veilings as an indicator of disk
evolutionary phases, since veilings are suggested to be correlated with
mass accretion rate. 
\citet{Kwan2007} used a comparison of their theoretical line profiles
with observed profiles from \citet{Edwards2006}.
They identified 11 and 15 CTTSs having blueshifted absorption features 
that resemble
the disk wind models and stellar wind models, respectively.
A summary of the statistics of the \ion{He}{1} absorption features for
CTTSs is presented in Table~\ref{tab:stat_HeI}. \citet{Edwards2006}
suggested that stars with high disk accretion rates are more likely to
have stellar wind signatures than disk wind signatures.
However, note that CTTSs in all veiling groups show morphological
features of group PC, with broad with broad redshifted absorption
suggesting stellar wind, even with the lowest observed veiling,
$\gamma_Y \le 0.2$.
In addition, all groups show profiles with narrow absorption features
suggesting disk wind, even with the highest observed veiling, $\gamma_Y
\ge 0.5$ (see Fig.~10 in \citealt{Edwards2006}).
This set of observations suggests that the trend is not so significant.
We find this unsurprising, since the veiling, and thus the mass
accretion rate, will not necessarily be a clear indicator of disk
evolution as discussed in Section~\ref{sec:targets}.
As a result, a clear progression of profile morphologies with veilings
is not observed.

\subsection{Chromospheric activities in Phase III sources}
\label{sec:chrom_III} 

\begin{figure*}[h]
\includegraphics[scale=0.3]{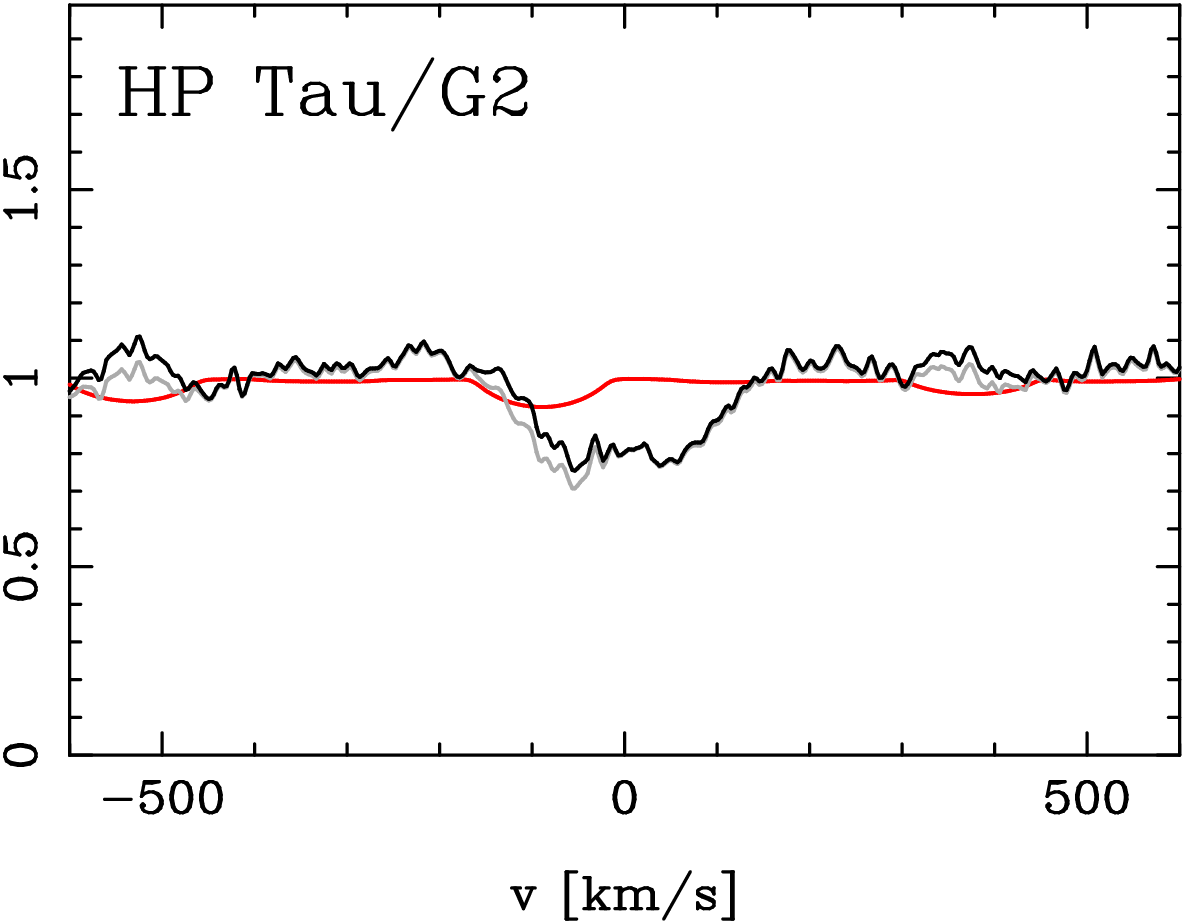}

   \vspace{1em}
\includegraphics[scale=0.3]{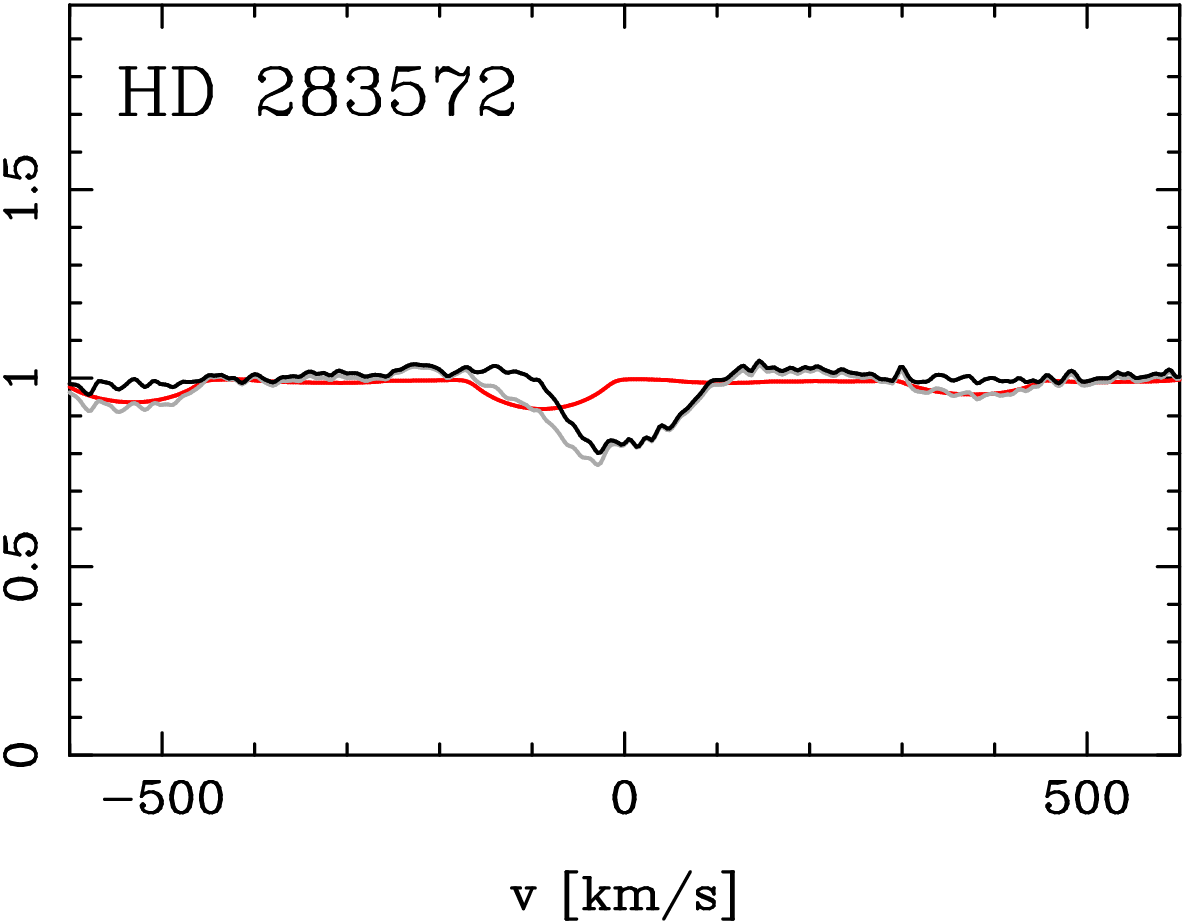}
   
   \vspace{1em}
\includegraphics[scale=0.3]{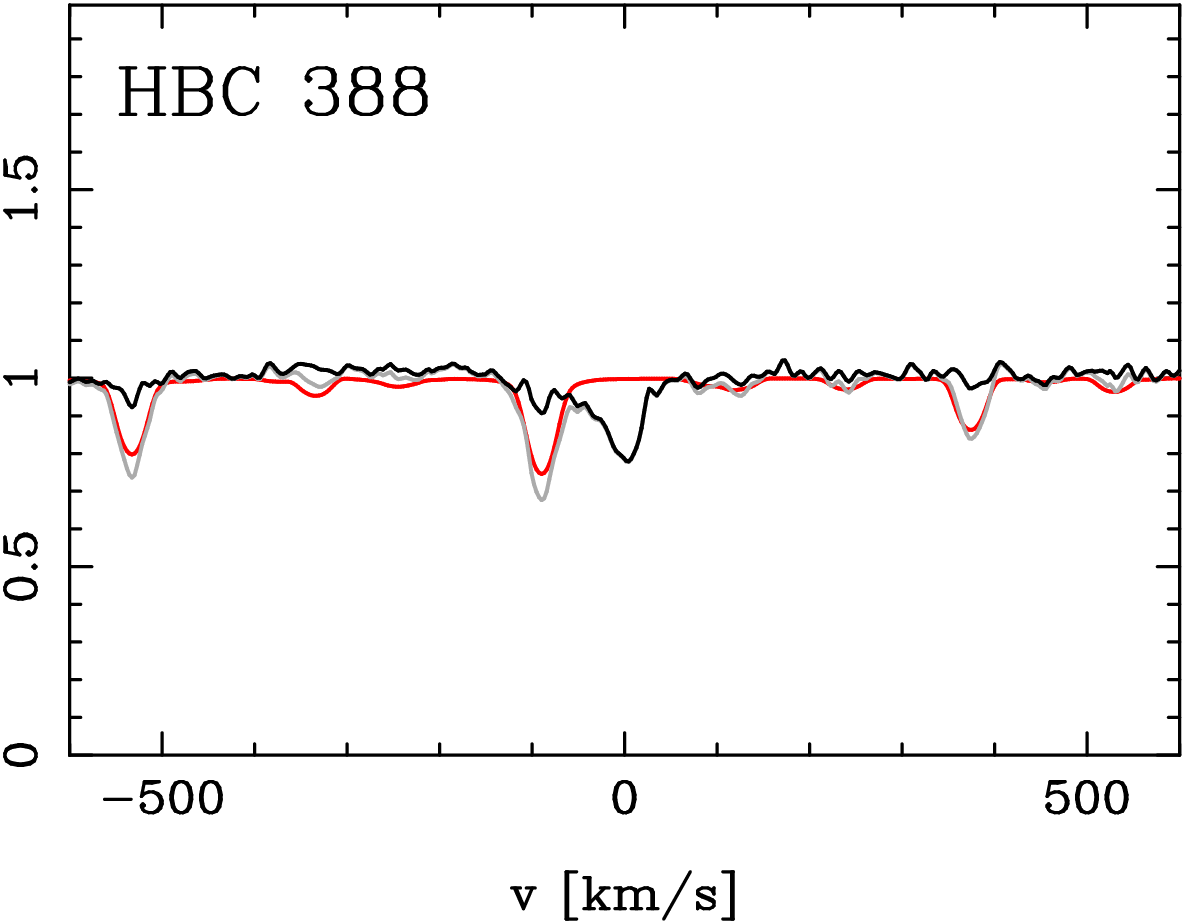}
   
  \vspace{1em}
\includegraphics[scale=0.3]{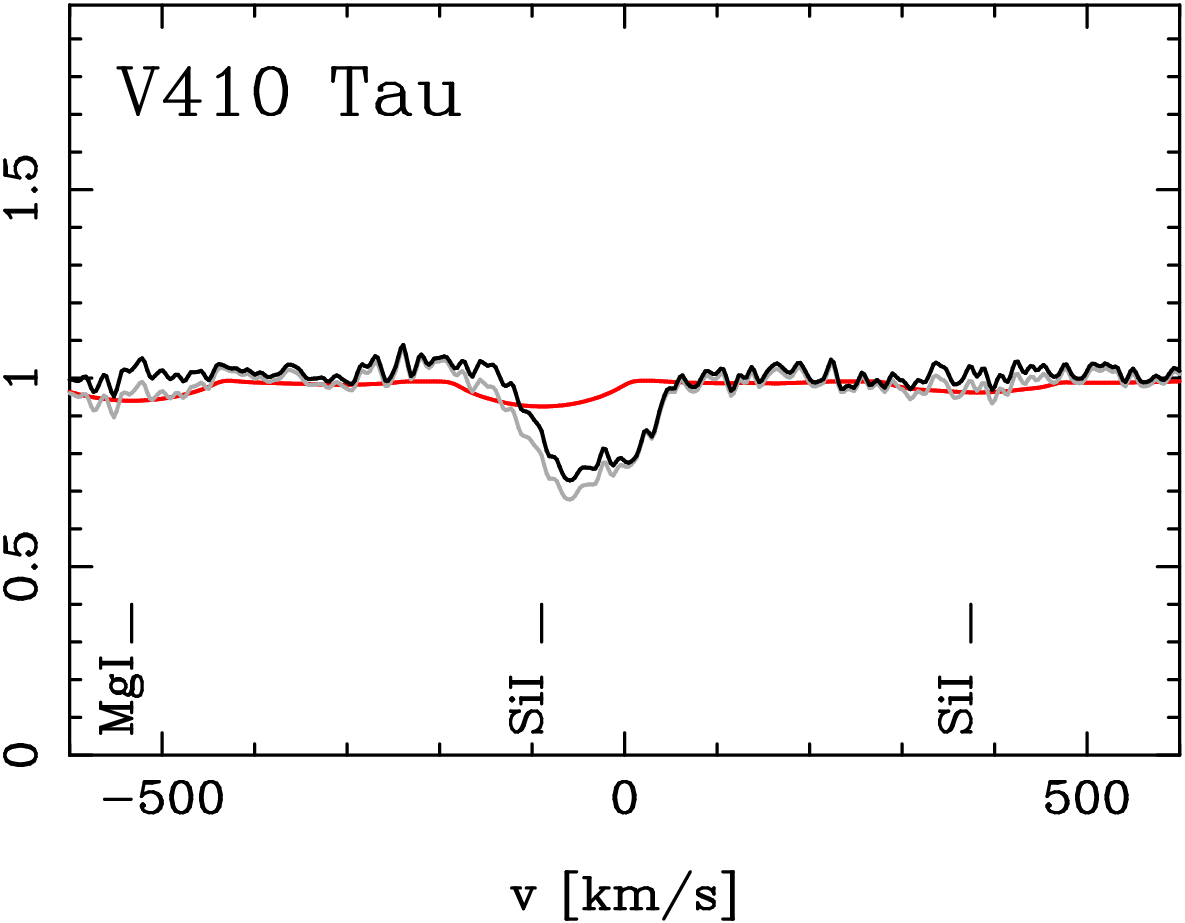}
  \caption{\ion{He}{1} $\lambda$10830 spectra for Phase III sources.
  The synthetic spectra, observed spectra, and residual profiles are
  shown with red, gray, and black lines, respectively.
Photospheric features, \ion{Mg}{1} $\lambda$10814.1 and \ion{Si}{1}
($\lambda$10830.1 and $\lambda$10846.8), are marked with vertical lines
   in the panel of V410 Tau (bottom).}
\label{fig:Synthe_PhaseIII}
\end{figure*}

In Section~\ref{sec:PhaseIII}, Phase III sources were suggested to show
no activities for mass accretion or inner winds. This is consistent with
WTTSs reported in \citet{Edwards2006}.
WTTSs are known to exhibit high levels of chromospheric and coronal
features \citep{Stelzer2013}, which persist at the age of
$\sim$50 to 100\,Myr \citep{Stauffer1994}.
The chromospheric profile has often been observed as emission lines in
X-ray and optical lines such as H$\alpha$, H$\beta$, \ion{Ca}{2}, and
\ion{Na}{1} lines.
Chromospheric activities are also recognized from \ion{He}{1}
$\lambda$10830 absorption lines (e.g., \citealt{Vaughan1968}), 
and are suggested to persist for star ages as advanced as those of Pop
II stars \citep{Takeda2011}.
We suggest that \HeI absorption features detected here are most likely
to be chromospheric features, considering that Phase III sources are
almost comparable to WTTSs \citep{Yasui2014} and that the observed \HeI
features in this paper are suggested to be stellar intrinsic features
(see Section~\ref{sec:PhaseIII}).
Note that the centered subcontinuum absorption features are not
necessarily indicative of the chromospheric activities.
When the features are accompanied by emissions, the profiles are
categorized into group DP (e.g., IRAS 04101+3103).
\citet{Cauley2014} suggested that the profiles are most likely to be
formed because of Keplerian rotation very close to the stellar surface.
Also, it should be noted that the photospheric features have little
impact on the \ion{He}{1} residual
profiles. Fig.~\ref{fig:Synthe_PhaseIII} shows synthetic spectra,
observed spectra, and residual profiles for all four Phase III sources.
Three photospheric features are observed in the spectral region,
\ion{Mg}{1} $\lambda$10814.1 and \ion{Si}{1} ($\lambda$10830.1 and
$\lambda$10846.8);
among them, only \ion{Si}{1} $\lambda$10830.1 profiles appear to cover
\ion{He}{1} profile. However, the line is very shallow for HP Tau/G2, HD
283572, and V410 Tau due to the large $v_{\rm broad}$,
$\sim$100\,km\,s$^{-1}$, while the line barely covers the \ion{He}{1}
profile for HBC 388 due to the small $v_{\rm broad}$, 25\,km\,s$^{-1}$.
Fig.~\ref{fig:Synthe_PhaseIII} suggests that the photospheric features
are moderately removed by subtracting synthetic spectra from observed
spectra (Section~\ref{sec:veil_PsF}).

\begin{figure*}[h]
\includegraphics[scale=0.5]{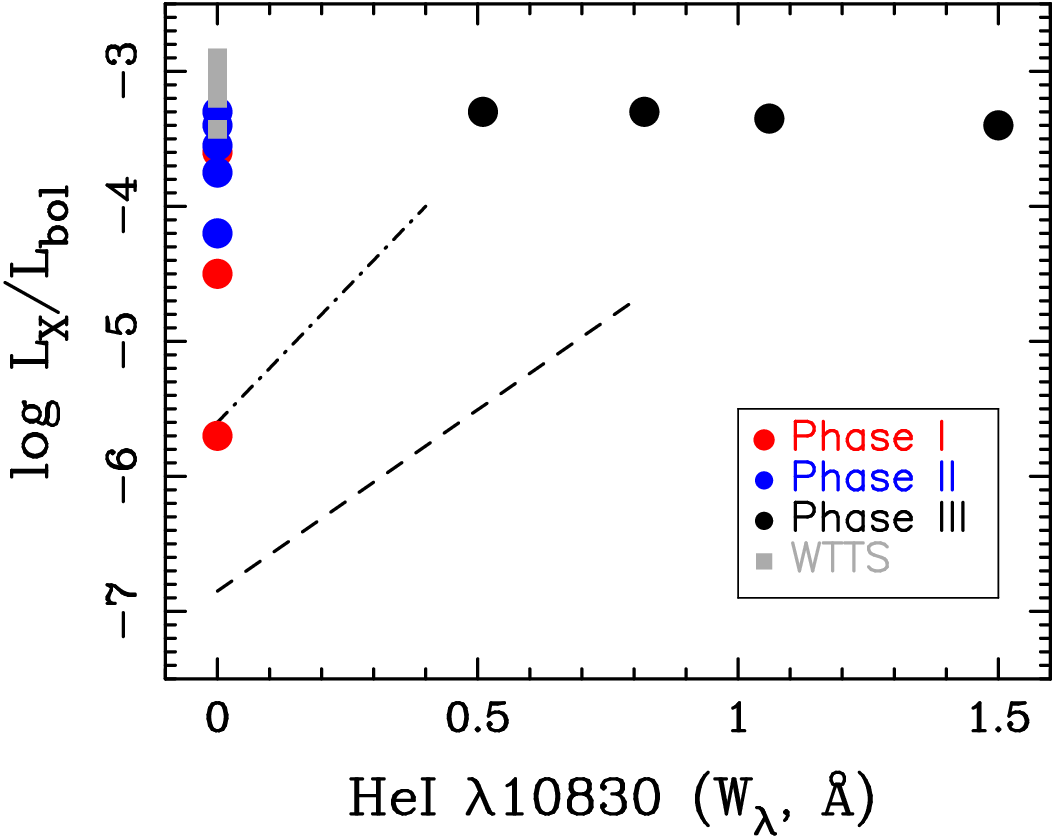}
\caption{The fractional X-ray luminosities ($L_{\rm X} / L_{\rm bol}$)
vs. \ion{He}{1} absorption equivalent widths.
Phase I, II, and III sources are shown by red, blue, and black circles,
respectively, while WTTSs in a sample of \citet{Edwards2006} are shown
by gray squares.
The lines show linear fits to sample stars in \citet{Zarro1986} by
\citet{Sanz-Forcada2008}; the dot-dashed line is for dwarfs and
subgiants, and the dashed line is for all giants.}  \label{fig:EWHeI_Lx}
\end{figure*}

The \ion{He}{1} line is a transition between lower metastable level of
2$^3$S and the upper level of 2$^3$P. Two mechanism have been proposed
for populating the upper levels of He I: photoionization-recombination
(PR) mechanism via EUV and X-ray radiations \citep{Goldberg1939} and
electron collisions from the ground level \citep{Andretta1997}.
The existence or nonexistence of correlation between \ion{He}{1}
absorption and X-ray luminosity can discern which process is dominant
\citep{Zarro1986}; the correlation exists when PR mechanism is dominant,
while the correlation is absent when collisional excitation is dominant.
We show fractional X-ray luminosities ($L_{\rm X} / L_{\rm bol}$) vs. 
\ion{He}{1} absorption equivalent widths (EWs) for all Phase III sources 
in Fig.~\ref{fig:EWHeI_Lx}. 
EWs are obtained from Table~\ref{tab:profile_HeI}, whereas $L_{\rm X} /
L_{\rm bol}$ is obtained from previous studies (\citealt{Gudel2007},
\citealt{Wichmann1996}).
All Phase III sources have $L_{\rm X} / L_{\rm bol}$ in the very small
range of $-$3.4--$-$3.5, whereas EWs span within $\sim$0.5--1.5.
For comparison, we also show the results from
\citet{Zarro1986} in Fig.~\ref{fig:EWHeI_Lx}, who presented the 
correlation in dwarfs and giants.
Thus, EW and $L_{\rm X} / L_{\rm bol}$ are not correlated in Phase III
sources, suggesting that the dominant mechanism for these sources is 
electron collisions rather than PR mechanism.
Phase I and II sources are plotted in Fig.~\ref{fig:EWHeI_Lx} for 
reference although we could not find any literature references for the
value of IRAS 04101+3103.

The \ion{He}{1} absorption features associated with chromospheric
activities had not been previously reported for YSOs, to our knowledge.
In \citet{Edwards2006}, no WTTSs show \HeI absorption features although
\HeI line profiles for two WTTSs show emission features (see
Section~\ref{sec:HeI}) and these authors surmised that these features
arise in active chromosphere or very weak accretion.
We suggest that the features detected even in such young phases are
characteristics of intermediate-mass stars, considering that these
authors' targets were primarily low-mass stars and that all of our
intermediate-mass star targets in Phase III show the \ion{He}{1}
absorption features. 
This pattern may be due solely to a positive correlation between stellar
mass and X-ray luminosities \citep{Gudel2007}, which can be correlated
with chromospheric activities.
We show $L_{\rm X} / L_{\rm bol}$ values of the WTTSs in
\citet{Edwards2006} in Fig.~\ref{fig:EWHeI_Lx} (gray squares).  The
$L_{\rm X} / L_{\rm bol}$ values for the WTTSs ($-$3.4--$-$2.9;
\citealt{Gudel2007}, \citealt{Bertout2007}, \citealt{Yang2012},
\citealt{Kastner2016}, \citealt{Kastner2004}), which are comparable to
those of our Phase III sources.
Therefore, the reason for the detection of \ion{He}{1} chromospheric
absorption features only in our intermediate-mass targets in Phase III
is not likely the correlation between stellar mass and X-ray
luminosities.

Correlations of chromospheric and higher-temperature features with
stellar age, magnetic field, and rotation period properties
have been previously reported \citep{Linsky2017}.
For the first factor, the \ion{He}{1} absorption features are not seen
in the sample WTTSs of \citet{Edwards2006}, which are also the sources
in the same star forming-region as our targets, i.e., the Taurus
star-forming region.
Therefore, the stellar age does not seem to be the cause for the
\ion{He}{1} detection in Phase III sources.
For the second factor, although there are only limited number of
derivation of magnetic fields, there seem to be no significant
differences between those for our Phase III sources and WTTSs reported
in \citet{Edwards2006}: $\sim$0.5--1.5\,kG for HBC 388 and V410 Tau
(\citealt{Basri1992}, \citealt{Skelly2010}), $\sim$0.5--2\,kG for LkCa
4, V891 Tau, and V827 Tau (\citealt{Donati2014}, \citealt{Donati2015},
\citealt{Giardino2006}).
Therefore, the magnetic field may not likely to be the cause
for the \ion{He}{1} detection in Phase III sources.
For the last factor, i.e., rotation, Phase III sources generally have
large $v \sin i$, $\sim$100\,km\,s$^{-1}$ (Table~\ref{tab:prop_W}),
while WTTSs in the sample of \citet{Edwards2006} generally have small
$v \sin i$ values, $<$20\,km\,s$^{-1}$ \citep{Rebull2004}.
Most theoretical models of stellar magnetic dynamos predict increasing
magnetic flux emerging through stellar photospheres with larger rotation
velocities \citep{Linsky2017}. This suggests that the larger rotation
velocities may be the possible cause for the detection of He I
absorptions in Phase III sources.


\subsection{Implication to theories of mass flow processes}
\label{sec:Implication}

Although configurations of mass flow processes around central stars have
been proposed, we do not yet have a conclusive theory explaining angular
momentum transport in the star formation process.
From the suggested progression of dominant mass flow processes with
evolution of protoplanetary disks in Section~\ref{sec:Progression}, we
discuss what factors determine which processes dominate, or in which
physical conditions one dominates the other.
These connections may help put useful constraints on current theories.

Before launching into this discussion, it is necessary to pay attention
to whether intermediate-mass stars can be regarded as simply massive T
Tauri stars.
\citet{Cauley2014} presented high-resolution \ion{He}{1} observations of
HAeBes, which correspond to intermediate-mass stars in Phase I. They
suggested that HAe stars, whose mass range is roughly comparable to that
of our targets, cannot be regarded as massive CTTSs
because they do not show signatures of magnetospheric accretion and
inner winds in the same manner as CTTSs: HAe stars show signatures of
magnetospheric accretion and stellar winds but do not show signatures of
disk winds\footnote{\citet{Cauley2014} suggested that HBe stars, in
general, do not show signatures of magnetospheric accretion and disk
winds, whereas show signatures of stellar winds.}.
They pointed out the possibility that the magnetospheres might be
smaller, based on the paucity of detected magnetic fields for HAeBes,
compared to CTTSs.
They confirmed small magnetospheres from their results that maximum
absorption velocities in the redshifted absorption profiles in their
sample are a smaller fraction of the system escape velocity than in
CTTSs.
They suggested that the small magnetospheres may be less efficient at
driving disk winds compared to CTTS magnetospheres.
The similar trends are also suggested for CTTSs by \citet{Kwan2007}:
stars with the highest disk accretion rates are less likely to have
redshifted absorption from magnetospheric funnel flows.
They suggested that their disks have small magnetospheres, which may set
up favorable conditions for stellar wind rather than disk wind. 
The maximum redshifted absorption velocities in all obtained \ion{He}{1}
line profiles of both Phase I and II sources in our intermediate-mass
targets having redshifted absorption features are estimated to be as
small as those of HAe stars, $\le$400\,km\,s$^{-1}$.
Assuming stellar radii of our targets (stellar mass of
$\gtrsim$1.5\,$M_\odot$) based on the isochrone model at the age of
1.5\,Myr by \citet{Siess2000}, escape velocities of our targets are
calculated to be $\gtrsim$\,500km\,s$^{-1}$.
The fraction of the system escape velocity estimated with the above
velocities (i.e., the maximum redshifted absorption velocities and
escape velocities) for our targets is smaller than for CTTSs,
corresponding to ballistic infall from distances of $<$2\,$R_\odot$
toward central stars.
This is again consistent with that for HAe stars (see Fig.~5 in
\citealt{Cauley2014}).
This suggests that magnetospheres of intermediate-mass stars are
generally smaller than those of CTTSs.
Nonetheless, signatures of disk winds are seen in Phase II sources
although the same trends (signatures of stellar winds) as seen in HAe
stars are seen in Phase I sources.
This suggests that the lack of disk wind signatures in HAe stars is not
a characteristic for intermediate-mass stars, but only for HAe stars, or
Phase I sources.
Further, there should be another factor that functions more effectively
than magnetosphere size for determining dominant mass flow processes.

The inner part of a protoplanetary disk in Phase II has low opacity and
is where most dust settling/growth is expected to be occurring.
In contrast, the inner disk in Phase I has high opacity but without
substantial dust settling/growth \citep{Yasui2014}.
Due to the low opacity in Phase II, radiation from central stars can
penetrate deeply into disks, and thus the ionization fraction should be
higher. This may be the reason why disk winds are active in Phase II.
This interpretation is consistent with the theory of disk wind in which
a sufficient magnetic field and ionization fraction launch
magnetocentrifugal winds over a significant range of radii in the disk,
from an inner truncation radius out to several AU \citep{Ouyed1997}.
In addition, Phase III stars, having already lost their entire dust
disks, show profiles with no signature of accretion or of inner winds.
The absence of these signatures in Phase III, along with the presence of
these signatures in Phase I and II, is consistent with the previous
suggestion that inner winds (stellar winds and disk winds) are
accretion-driven.

\section{Conclusion} \label{sec:conc}



We performed near-infrared high-resolution ($R = 28$,000) spectroscopy
of {13} young intermediate-mass stars in the Taurus star-forming region
with WINERED.
Our obtained spectra ($\lambda = 0.91$--1.35\,$\mu$m) depict \HeI
$\lambda$10830 and P$\beta$ lines that are sensitive to magnetospheric
accretion and winds.
We investigate five sources each for P$\beta$ and \ion{He}{1} lines from
previous studies, in addition to the 13 targets observed in this study.
Based on the presence of near- and mid-infrared continuum emission,
young intermediate-mass stars can be classified into three different
evolutionary stages: Phases I, II, and III in the order of evolution.

\begin{enumerate} 
 \item Phase I and II sources exhibit a variety of profile morphologies in
       P$\beta$, while all Phase III sources are in group F (i.e., 
       they are featureless). 
       The statistics of the P$\beta$ profile morphologies for Phase I 
      and Phase II sources are not significantly different except that
      almost half of Phase I sources show group E features, 
      while none of Phase II sources exhibit the features.
      The statistics of Phase I and II sources are not inconsistent 
       with those of low-mass CTTSs.
      Furthermore, the morphology statistics for Phase III sources
      are consistent with those of low-mass WTTSs.

 \item Phase I and II sources also show variety of \ion{He}{1} 
	 profile morphologies. The profile morphologies of Phase I and Phase 
	 II sources are mostly broad subcontinuum absorption lines and 
	 narrow subcontinuum absorption lines, respectively; 
	 however, the statistics of these profile morphologies are not 
	 significntly different and are comparable to those of low-mass
	 CTTSs. 
       On the contrary, all Phase III sources are in group A (i.e., they exhibit
       pure absorption) with centered subcontinuum absorption features, 
       inconsistent with the characteristics of low-mass WTTSs.

\item By comparing observed P$\beta$ and \ion{He}{1} line profiles with
      model profiles,
      dominant mass flow processes are suggested to be stellar wind and
      probably magnetospheric accretion in Phase I; magnetospheric
      accretion and disk wind in Phase II; and no activities in Phase
      III.

\item The different dominant processes in different phases indicate a
      clear progression of dominant mass flow processes with disk
      evolution.
      However, because this study may not contain a statistically
      significant number of sources, more statistical studies are
      necessary to draw a firm conclusion.

\item The mass flow processes in HAe stars are consistent with Phase I,
      with stellar wind as the dominant process.
      As for the progression of mass flow processes, the trends in
      veilings found by other authors do not seem so significant
      compared to that suggested in this paper based on disk
      evolutionary phase.
      Although the absence of activities in Phase III is consistent with
      previous studies for low-mass WTTSs,
      we note that \ion{He}{1} line profiles do show chromospheric
      activities in these sources.

\item Disk wind signatures are seen in Phase II sources, despite their
      smaller magnetospheres, which was suggested to be the reason for
      the absence of the signature in HAe stars by other authors.
      Alternatively, we suggest that opacity in protoplanetary disks
      plays an important role in determining dominant mass flow 
      processes.

\end{enumerate}

\acknowledgments

We thank an anonymous referee for helpful comments that improved the
manuscript.
%
This work has made use of the VALD database, operated at Uppsala
University, the Institute of Astronomy RAS in Moscow, and the University
of Vienna.

\begin{table*}
\caption{Target list and photometric data. } \label{tab:target}
\begin{center}
\begin{tabular}{lllllllllll}
\hline
\hline
 Object & R.A. & Decl. & SpT & $J$ & $H$
 & $K_S$ & $\alpha_{\rm MIR}$\\
& (J2000.0) & (J2000.0) & & (mag) & (mag) & (mag) \\
 (1) & (2) & (3) & (4) & (5) & (6) & (7) & (8) \\ 
\hline
V892 Tau & 04:18:40.62 & +28:19:15.5 & B9 & 8.5 & 6.6 & 6.2 & 1.75 \\
 AB Aur & 04:55:45.85 & +30:33:04.3 & A0 & 6.3* & 5.5* & 4.5* & $-$0.79 \\
IRAS 04101+3103	& 04:13:20.02 & +31:10:47.3 & A1 & 9.2 & 8.7 & 8.1 & 0.97 \\
HP Tau/G2 & 04:35:54.15 & +22:54:13.6 & G0 & 8.1 & 7.5 & 7.2 & $-$0.43 \\
RY Tau & 04:21:57.41 & +28:26:35.6 & G1 & 7.2 & 6.1 & 5.4 & $-$0.16 \\
SU Aur & 04:55:59.38 & +30:34:01.5 & G1 & 7.2 & 6.6 & 6.0 & $-$0.32 \\
HD 283572 & 04:21:58.84 & +28:18:06.5 & G5 & 7.4 & 7.0 & 6.9 & $-$2.85 \\
T Tau  & 04:21:59.43 & +19:32:06.4 & K0 & 7.2 & 6.2 & 5.3 & $-$0.56 \\
HBC 388  & 04:27:10.57 & +17:50:42.6 & K1 & 8.8 & 8.4 & 8.3 & $-$2.78 \\
RW Aur & 05:07:49.57 & +30:24:05.2 & K3 & 8.4 & 7.6 & 7.0 & $-$0.67 \\
V773 Tau & 04:14:12.92 & +28:12:12.3 & K3 & 7.5 & 6.6 & 6.2 & $-$1.10 \\
V410 Tau & 04:18:31.11 & +28:27:16.1 & K3 & 8.4 & 7.8 & 7.6 & $-$2.78 \\
UX Tau  & 04:30:03.99 & +18:13:49.4 & K5 & 8.6 &  8.0 & 7.6 & $-$1.99 \\
\hline
\end{tabular}
\end{center}
 {{\bf Notes.}\\
Col. (4) Spectral types from \citet{{Furlan2006}, {Furlan2011}}.
Cols. (5)--(7) NIR magnitude from the 2MASS Point Source Catalog, {\it
J}-band (Col. 5), {\it H}-band (Col. 6), and $K_S$-band magnitude
(Col. 7).
Magnitudes for AB Aur are obtained from a study reported by
\citet{Kenyon1995}, shown with ‘*’.
Col. (8) SED slope ($\alpha = d \ln \lambda F \lambda / d
\ln \lambda$) with Spitzer IRS 6--13\,$\mu$m presented in
\citet{Furlan2006}.}
\end{table*}

\begin{table*}
\caption{Disk properties of target YSOs.}
\label{tab:disk_prop}
\begin{center}
\fontsize{6pt}{0pt}\selectfont
\begin{tabular}{lllllllllllll}
\hline
\hline
Object 
	& Phase 
	& SED 
	& C/WTTS
	& $i$ 
	& $\dot{M}$ 
	& $M_{\rm disk}$ \\
 & & & & (deg.) & ($10^{-8}$\,$M_\odot$\,yr$^{-1}$)
& ($M_\odot$) \\
(1) & (2) & (3) & (4) & (5) & (6) & (7)\\
\hline
 V892 Tau & I & II & W & 60 (Mo08), {59 (Ha10)}
& 10.5 (DNB18), 7.2 (Liu11) & 0.009 \\
AB Aur & I & II & C & $<$45 (Gr99), 22 (Co05), 40 (Ta12)  
& 14.1 (GNTH06), 1.8 (DB11), 39 (DNB18) & 0.004 \\
IRAS 04101+3103 & I & II & C & ... & ... & ... \\
HP Tau/G2 & III & III & W & 50 (We87), 67 (Bo95) & ... & ... \\
RY Tau & II & II & C & 86 (Mu03), 66 (Is10) & 6.4--9.2 (Ca04)
& 0.02 \\
SU Aur & II & II & C & 62 (Ak02), 86 (Mu03) & 0.5--0.6 (Ca04)
& 0.0009 \\
HD 283572 & III & III & W &  35 (St98), 48 (Wi02), 60 (JBB94) 
& ... & $<$0.0004 \\
T Tau & I & II & C & 19 (HRB97), 20 (Ar02)
& 3.1--5.7 (Ca04), 3.2 (WG01) & 0.008 \\
 HBC 388 & III & III & W &  45 (Ar02)
& $<$0.9 (WG01) & $<$0.0003 \\
RW Aur & II & II & C &  40 (Ar02) & 2.0 (In13) & 0.004 \\
V773 Tau & II & II & W & 34 (Si16) & 0.15 (Si16) & 0.0005 \\
V410 Tau & III & III & W & 54 (Ha95), 70 (JBM94, St94, RS96),
& $<$0.16 (WG01) & $<$0.0004\\
& & & &
80 (He89), 90 (Bo95) & \\
UX Tau & II & II & C & 35 (An11) & 1.1 (Es10) & 0.005 \\
\hline
\end{tabular}
\end{center}
{{\bf Notes.}\\
Col. (2) Phase of dust disk (Phase I/II/III) from \citet{Yasui2014} (see
 Section~\ref{sec:targets}). 
Col. (3) SED classification type (Class I/II/III). 
Col. (4) Accretion-based classification type (CTTS or WTTS). 
 Col. (5) Inclination angles.
Note that the angles for Phase III sources are often estimated with
stellar rotation periods and stellar rotational velocities, while those
for Phase I and II sources are estimated from observations of their
disks.
Col. (6) Mass accretion rate.  
Col. (7) Dust disk masses derived by \citet{Andrews2005}. \\
 {\bf References.} \\
Mo08: \citet{Monnier2008}, 
Ha10: \citet{Hamidouche2010}, 
DNB18: \citet{Dong2018}, 
Liu11: \citealt{Liu2011}, 
Gr99: \citet{Grady1999}, 
Co05: \citet{Corder2005}, 
Ta12: \citet{Tang2012}, 
GNTH06: \citet{Garcia Lopez2006},
DB11: \citet{Donehew2011},
We87: \citet{Weaver1987}, 
Bo95: \citet{Bouvier1995}, 
Mu03: \citet{Muzerolle2003},
Is10: \citet{Isella2010},
Ca04: \citet{Calvet2004}, 
Ak02: \citet{Akeson2002},
St98: \citet{Strassmeier1998},
Wi02: \citet{Wittkowski2002}, 
JBB94: \citet{Joncour1994a}, 
HRB97: \citet{Herbst1997}, 
Ar02: \citet{Ardila2002},
WG01: \citet{White2001},
In13: \citet{Ingleby2013},
Si16: \citet{Simon2016}, 
Ha95: \citet{Hatzes1995}, 
JBM94: \citet{Joncour1994b}, 
St94: \citet{Strassmeier1994}, 
RS96: \citet{Rice1996}}, 
He89: \citet{Herbst1989},
An11: \citet{Andrews2011},
Es10: \citet{Espaillat2010}. 
\end{table*}

\begin{table*}
 \caption{Intermediate-mass stars in the samples of previous studies
\citep{{Folha2001}, {Edwards2006}} and their properties.}
\label{tab:target_others}
\begin{center}
\begin{tabular}{llllllllllllll}
\hline
\hline
 Object & SpT & Phase
& \multicolumn{3}{c}{P$\beta$ (P$\gamma$)}
& \multicolumn{3}{c}{\ion{He}{1}} \\
& & & \multicolumn{3}{c}{------------------------------------------}
  & \multicolumn{3}{c}{------------------------------} \\
 & & &  Data & Group & Abs. Type & Data & Group & Abs. Type \\
 (1) & (2) & (3) & (4) & (5) & (6) & (7) & (8) & (9) \\
\hline
CW Tau & K3 & I & \checkmark  (\checkmark) & DP (E) & ... (...)
& \checkmark & DP & ... \\
HP Tau & K3 & I & \checkmark & IPC & r \\
GM Aur & K3 & II & \checkmark (\checkmark )& IPC (F) & r  (...)
& \checkmark & F & ... \\
DR Tau & K5 & I & \checkmark (\checkmark) & DP (IPC) & ...  (r)
& \checkmark & PC & b \\
DS Tau & K5 & II & \checkmark (\checkmark) & IPC (IPC) & r  (r)
& \checkmark & BR & b, r \\
HN Tau & K5 & I & \multicolumn{1}{r}{(\checkmark)} & (E) & 
\ \ (...) 
& \checkmark & DP & ... \\
\hline
\end{tabular}
\end{center}
{{\bf Notes.}\\
Col. (2): Spectral types from \citet{{Furlan2006},{Furlan2011}}.
Col. (3): Phase of dust disk (Phase I/II/III) from \citet{Yasui2014}
 (see Section~\ref{sec:targets}). 
 Cols. (4)--(6): Properties of the P$\beta$ line profiles {with those of
 P$\gamma$ profiles (shown in parenthes)}.
 The tick marks in Col. 4 mean that {the corresponding} spectra are
  available in \citet{Folha2001} {for P$\beta$ and \citet{Edwards2006}
  for P$\gamma$}.
Col. (5): Morphology groups based on the classification scheme in 
 Section~\ref{sec:Pbeta}.
Col. (6): Type of subcontinuum absorption (blue, red, or centered
subcontinuum absorption, denoted by b, r, and c, respectively).
Cols. (7)--(9): Properties of \ion{He}{1} line profiles; {the tick}
marks in Col. 7 {indicate} that {the corresponding} spectra are
available in \citet{Edwards2006}.
Col. (8): Morphology groups based on the classification scheme in
Section~\ref{sec:Pbeta}.
Col. (9): Type of subcontinuum absorption (blue, red, or centered
subcontinuum absorption{, denoted} by b, r, and c, respectively).}
\end{table*}

\begin{table}
\caption{Summary of WINERED Observations.} \label{tab:obs}
\begin{center}
\begin{tabular}{lllllllllll}
\hline
\hline
Object
 & Obs date 
 & Airmass & Exp time & S/N$^a$ & Standard$^b$ \\
 & (YYYY/MM/DD)
 & (sec z)    & (s) 
 & \\
\hline
V892 Tau
     & 2013/12/08 & 1.0--1.1 & 4800 ($600 \times 8$)
     & 58 & HR 104 (A2Vs) \\
AB Aur & 2013/02/22
	& 1.2	& 1200 $(300 \times 4)$ 
	& 82 & HIP53910 (A1V) \\
IRAS 04101+3103	& 2014/09/28 
	& 1.1--1.2 & 3000 ($300 \times 10$) 
	& 59 & HR 196 (A2Vs) \\
HP Tau/G2& 2013/12/04
	& 1.0 &	2400 ($600 \times 4$) 
	& 54  & HD 1041 (M1III) \\
RY Tau	& 2013/02/23 
	& 1.3--1.4 & 1200 ($300 \times 4$) 
	& 43  & HIP58001 (A0Ve+K2V) \\
	& 2013/02/23 
	& 1.5--1.8 & 1200 ($300 \times 4$) 
	& & HIP23179 (A1V) \\
SU Aur & 2013/03/03 
	& 1.8-2.5 & 1800 ($300 \times 6$) 
	& 34 &  HIP76267 (A0V) \\
HD 283572 & 2013/11/30 & 1.0 & 2400 ($600 \times 4$) 
& 113  & omi Aur (A2Vp) \\
T Tau & 2013/12/02 & 1.2-1.3 &	7200 ($1200 \times 6$) 
& 94 & HR 196 (A2Vs) \\
HBC 388 & 2014/10/15 
	& 1.2-1.3 & 2400 ($600 \times 4$) 
	& 84  & HR 922 (B9V) \\
RW Aur 	& 2013/12/03
	& 1.0 &	4800 ($600 \times 8$) 
	& 55  & HR1736 (A2V) \\
V773 Tau & 2013/11/28 & 1.0-1.1 & 5400 ($600 \times 9$) 
& 74 & 50 Cas (A2V) \\
V410 Tau & 2013/12/08 & 1.1--1.3 &	3600 ($600 \times 6$)
&  33 & HR 104 (A2Vs) \\
UX Tau & 2013/12/12 & 1.3 & 7200 ($600 \times 12$) 
	& 37  &  50 Cas (A2V) \\
\hline
\end{tabular}
\end{center}
{{\bf Notes.}\\
$^a$The pixel-to-pixel S/N of the continuum level at $\lambda \sim
12700$\,\AA. For RY Tau, the value (43) is from combined spectra of all
8 frames. \\ 
$^b$Standard stars used for telluric correction. The spectral type for
each standard star is shown in parentheses.}
\end{table}


\begin{table*}[!h]
\caption{Obtained properties from the fit of the synthetic spectra.} 
\label{tab:prop_W}
\begin{center}
\begin{tabular}{lllllllllllll}
\hline
\hline
Object & SpT & $v_{\rm broad}$ 
& RV & $\gamma_Y$ & $\gamma_J$ & \\

 &  & (km s$^{-1}$) & (km s$^{-1}$) & \\

(1) & (2) & (3) & (4) & (5) & (6) \\
\hline
V892 Tau & A0 (B9) & (100$^a$) & (16.0$^b$) & ... & ... \\ 
AB Aur & A0 (A0) & (116$^c$) & (24.7$^c$) &
		 ... & ... \\ 
		
IRAS 04101+3103 & A0 (A1) & (100) & (20) & ... & ... \\
HP Tau/G2 & G0 (G0) & 120 & 15.1 & 0.1 & 0.1 \\
RY Tau & G0 (G1) & 65 & 22.9 & 0.4 & 0.3 \\
SU Aur & G0 (G1) & 85 & 23.9 & 0.3 & 0.2 \\
HD 283572 & G5 (G5) & 105 & 17.9 & 0.0 & 0.1 \\
T Tau & K0 (K0) & 35 & 21.8 & 0.9 & 0.5 \\
HBC 388 & K0 (K1) & 25 & 15.7 & 0.0 & 0.0 \\
RW Aur & K3 (K3) & 35 & 13.5 & 1.0 & 0.2 \\
V773 Tau & K3 (K3) & 60 & 5.6 & 0.4 & 0.3 \\ 
V410 Tau & K3 (K3) & 130 & 26.2 & 0.0  & 0.0 \\
UX Tau & K5 (K5) & 25 & 12.5 & 0.6 & 1.0 \\
\hline
\end{tabular}
\end{center}
{{\bf Notes.}\\
Col. (2) Spectral types of target objects used for synthetic
spectra. Spectral types from the literature are shown in parentheses.
Col. (3) Line broadening measured in velocity units.
For V892 Tau and AB Aur, values from the literature are assumed, shown 
in parentheses.
For IRAS 04101+3103, 100\,km\,s$^{-1}$ is assumed, shown in 
parenthesis (see more detail in the main text). 
Col. (4) Radial velocities. For V892 Tau and AB Aur, values from
the literature are assumed.  For IRAS 04101+3103, the average RV value for
 all targets, 20\,km\,s$^{-1}$, is assumed (see more detail in the main
 text).
Col. (5) {\it Y}-band veiling. 
Col. (6) {\it J}-band veiling. \\ \\
{\bf References.}\\
$^a$\citet{Mooley2013}, 
$^b$\citet{Bertout2006}, 
$^c$\citet{Alecian2013}.} 
\end{table*}


\begin{table*}[!h]
 \caption{P$\beta$ profile parameters.} 
\label{tab:profile_Pbeta} 
\begin{center}
\fontsize{6.pt}{0pt}\selectfont
\begin{tabular}{lllllllllllll}
\hline
\hline
Object & Profile type 
 & $V^{\rm blue}_{\rm max}$ & $V^{\rm red}_{\rm max}$
 & $V^{\rm em}_{\rm peak}$ & $V^{\rm abs}_{\rm peak}$
 & $F^{\rm em}_{\rm peak}$ & $F^{\rm abs}_{\rm peak}$
 & Em. $W_\lambda$ & Ab. $W_\lambda$ 
 & Ab. Type \\
 &  & (km s$^{-1}$) & (km s$^{-1}$) 
 & (km s$^{-1}$) & (km s$^{-1}$) & 
 &  & (\AA) & (\AA) & \\
(1) & (2) & (3) & (4) & (5) & (6) & (7) & (8) & (9) & (10) & (11) \\
\hline
V892 Tau & E 
 & $-$350 ($-$250) & 250 (200)  & $-$20 & ... & 2.01 (1.64) & ... 
 & 8.93 (5.74) & ... & ... \\
AB Aur & E 
 & $-$400 ($-$350) & 400 (300)  & $-$20
 & ... & 3.39 (2.69) & ... & 19.0 (13.1) & ... & ... \\
IRAS 04101+3103 & E 
& $-$500 ($-$350)
& 400 (350) & 5 & ... & 1.98 (1.57) & ...
& 12.2 (7.2) & ... & ... \\
HP Tau/G2 & F 
 & ... & ... & ... & ... & ... & ... & ... & ... & ... \\
RY Tau & DP 
& $-$400 & 300 & $-$100 
& ... & 1.27 & ... & 4.00 & ... & ... \\
SU Aur & DP 
 & $-$300 & 250 & $-$100 & ... & 1.22 & ... & 2.29 & ... & ... \\
HD 283572 & F 
 & ... & ... & ... & ... & ... & ... & ... & ... & ... \\
T Tau & E & $-$250 & 250 
& $-$15 & ... & 2.56 & ... & 9.15 & ... & ... \\
HBC 388 & F 
 & ... & ... & ... & ... & ... & ... & ... & ... & ... \\
RW Aur & IPC & $-$400 & 450 
& $-$50 & 305 & 1.99 & 0.96 & 10.36 & 0.20 & r \\
V773 Tau & F 
 & ... & ... & ... & ... & ... & ... & ... & ... & ... \\
V410 Tau & F 
 & ... & ... & ... & ... & ... & ... & ... & ... & ... \\
UX Tau & DP 
& $-$250 & 250 & $-$35 & ... & 1.29 & ... & 1.68 & ... & ... \\
\hline
\end{tabular}
\end{center}
{{\bf Notes.}\\
Col. (2) Profile type based on classification scheme in
Section~\ref{sec:Pbeta} (see main text in the detail).
Cols. (3) and (4): The maximum blueshifted and redshifted line 
velocities.
For V892 Tau, AB Aur, and IRAS 04101+3103, estimated values assuming
 $\gamma_Y = \gamma_J = 0.0$ are shown, while those assuming $\gamma_Y =
 2.0$, $\gamma_J = 2.0$ are shown in parentheses. 
Cols. (5) and (6): The velocity of peak emission and absorption. 
Cols. (7) and (8): The line fluxes relative to the continuum at the peak
emission and absorption velocities. 
Cols. (9) and (10): Emission and absorption equivalent widths. 
Col. (11): The type of subcontinuum absorption: blue,
red, or centered subcontinuum absorption are marked by the letter b, r,
and c, respectively.}
\end{table*}


\begin{table*}[!h]
\caption{\ion{He}{1} Profile parameters.}
\label{tab:profile_HeI} 
\begin{center}
\fontsize{6.pt}{0pt}\selectfont
\begin{tabular}{lllllllllllll}
\hline
\hline
Object & Profile type 
 & $V^{\rm blue}_{\rm max}$ & $V^{\rm red}_{\rm max}$
 & $V^{\rm em}_{\rm peak}$ & $V^{\rm abs}_{\rm peak}$
 & $F^{\rm em}_{\rm peak}$ & $F^{\rm abs}_{\rm peak}$
 & Em. $W$ & Ab. $W$ 
 & Ab. Type \\
 &  & (km s$^{-1}$) & (km s$^{-1}$) & (km s$^{-1}$) & (km s$^{-1}$) & 
 &  & (\AA) & (\AA) & \\
(1) & (2) & (3) & (4) & (5) & (6) & (7) & (8) & (9) & (10) & (11) \\
\hline
V892 Tau & IPC 
 & $-$250 & 200 & $-$155 & 80 & 1.19 & 0.88 & 0.18 & 0.09 & r \\
%
AB Aur & PC 
 & $-$400 & 400 & 45 & $-$225 & 1.29 & 0.79 & 2.71  & 0.98 & b \\
%
IRAS 04101+3103 & DP 
 & $-$450 & 350  & $-$230 & 0 & 1.16 & 0.85 & 1.12 & 0.48 & c \\ 
HP Tau/G2 & A 
 & $-$150 & 150 & ... & $\sim$0 & ... & 0.75 & ... & 1.50 & c \\
%
RY Tau & DP & $-$400 & 300 & $-$155 & $-$40
& 1.41 & 0.79 & 3.03 & 0.32 & b \\
SU Aur &  BR 
& $-$200 & 250 & ... & $-$95 & ... & 0.47 & ... & 2.32 & b, r \\
HD 283572 & A 
 & $-$100 & 100 & ... & $\sim$0 & ... & 0.83 & ... & 0.82 & c \\
T Tau & PC 
& $-300$ & 400 & $-$10  & $-$185  & 2.31 & 0.13 & 7.47 & 3.98 & b \\
HBC 388 & A 
& $-$100 & 50 & ... & 0 & ... & 0.78 & ... & 0.51 & c \\
RW Aur & IPC 
& $-$400 & 350 & $-$140 & 60 & 1.81 & 0.76 & 4.32 & 1.73 & r \\
V773 Tau & PC 
 & $-$100 & 250 & 30 & $-$15 & 1.49 & 0.70 & 1.32 & 0.44 & b \\
V410 Tau & A
 & $-$160 & 120 & ... & $\sim$0 & ... & 0.69 & ... & 1.06 & c \\
UX Tau & DP & $-$300 & 250 & $-$85 & 50
& 1.71 & 0.74 & 3.69 & 0.33 & r \\
\hline
\end{tabular}
\end{center}
{{\bf Notes.}\\ Col. (1): Object name.  Col. (2): Profile type based on
 classification scheme in Section~\ref{sec:Pbeta} (see main text in the
 detail).
Cols. (3) and (4): The maximum blueshifted and redshifted line
velocities.
Cols. (5) and (6): the velocity of peak emission and absorption. 
Cols. (7) and (8): The line fluxes relative to the continuum at the peak
emission and absorption velocities. 
Cols. (9) and (10): Emission and absorption equivalent widths. 
Col. (11): The type of subcontinuum absorption: blue, red, or centered
 subcontinuum absorption are marked by the letter b, r, and c,
 respectively.}
\end{table*}

\begin{table*}[!h]
 \caption{Statistics summary of P$\beta$ features.}
\label{tab:stat_Pbeta} 
\begin{center}
 \fontsize{6.pt}{0pt}\selectfont
\begin{tabular}{llllllllc}
\hline
\hline
	& \multicolumn{1}{c}{E}
	& \multicolumn{1}{c}{DP} 
	& \multicolumn{1}{c}{PC}
	& \multicolumn{1}{c}{IPC}
	& \multicolumn{1}{c}{A}
	& \multicolumn{1}{c}{F} 
	& \multicolumn{1}{c}{BR} 
	& \color{black} Reference \\
\hline  
\multicolumn{3}{l}{\bf Intermediate-mass stars} \\

 Phase I & \color{black} 57\,\% (4/7) & 29\,\% (2/7)
	 & 0\,\% (0/7) & 14\,\% (1/7)
	 & 0\,\% (0/7) &  0\,\% (0/7)
	 & 0\,\% (0/7) 
	 & This study\textcolor{black}{, FE01} \\
 
 Phase II & \color{black} 0\,\% (0/7) & \color{black} 43\,\% (3/7)
	 & \color{black} 0\,\% (0/7) & \color{black} 43\,\% (3/7)
	 & \color{black} 0\,\% (0/7) & \color{black} 14\,\% (1/7)
	 & 0\,\% (0/7) 
	 & This study, FE01 \\

 Phase III & 0\,\% (0/4) & 0\,\% (0/4) 
& 0\,\% (0/4) & 0\,\% (0/4) & 0\,\% (0/4) & 100\,\% (4/4)
	 & 0\,\% (0/4) & This study, FE01 \\
\hline
\multicolumn{3}{l}{\bf Low-mass stars} \\
CTTS$^a$
	& 49\,\% (20/41)
	& \color{black} 12\,\% (5/41) 
	& 0\,\% (0/41)
	& 32\,\% (13/41)
	& 0\,\% (0/41) & 7\,\% (3/41) 
	& 0\,\% (0/41) 
	& FE01 \\
WTTS
 & 0\,\% (0/3) & 0\,\% (0/3) 
	& 0\,\% (0/3) & 0\,\% (0/3)
	& 0\,\% (0/3) & 100\,\% (3/3) 
	& 0\,\% (0/3) 
	& FE01 \\
CTTS (P$\gamma$)$^b$
	& 69\,\% (27/39)
	& 3\,\% (1/39)
	& 0\,\% (0/39)
	& 21\,\% (8/39)
	& 0\,\% (0/39)
	& 8\,\% (3/39)
	& 0\,\% (0/39) 
	& Ed06 \\
WTTS (P$\gamma$)$^c$ 
	& 17\,\% (1/6)
	& 0\,\% (0/6) 
	& 0\,\% (0/6)
	& 0\,\% (0/6)
	& 0\,\% (0/6)
	& 83\,\% (5/6) 
	& 0\,\% (0/6) 
	& Ed06 \\
	\hline
\end{tabular}
\end{center}
{{\bf Notes.}\\ 
$^a$The line profiles for 41 CTTSs in a sample of
\citet{Folha2001} with sufficient S/Ns were used (see main text in
detail). \\
$^b$The P$\beta$ line profiles for all CTTSs except for 12 sources (DR
Tau, RW Aur A, DS Tau, YY Ori, AA Tau, GI Tau, CI Tau, BM And, RW Aur B,
SU Aur, UX Tau, and GM Aur) are categorized into
 group E. The line profiles for the 12
sources are as follows:
group DP: SU Aur;
group IPC: DR Tau, RW Aur A, DS Tau, YY Ori, AA Tau, GI Tau, BM And, and
 RW Aur B;
Group F: CI Tau, UX Tau, and GM Aur. \\
$^c$The P$\beta$ line profiles for all WTTSs except for V826 Tau are 
categorized into group F.
Although classification is difficult for V826 Tau due to very subtle
features ($W_\lambda < 0.3$\,\AA), we categorize it into group E. \\
{\bf References.} \\
FE01: \citet{Folha2001}, Ed06: \citet{Edwards2006}.}
\end{table*}

\begin{table*}[!h]
 \caption{
 Statistics summary of \HeI features {and 
 blueshifted absorption types.}}
  \label{tab:stat_HeI}
\begin{center}
\fontsize{6.pt}{0pt}\selectfont
\begin{tabular}{llllllll|ll|l}
\hline
 \hline
 	& \multicolumn{7}{c|}{Spectral type}
 	& \multicolumn{2}{c|}{Blue abs. type} 
	& \color{black} Reference \\
	
	& \multicolumn{1}{c}{E}
	& \multicolumn{1}{c}{DP} 
	& \multicolumn{1}{c}{PC}
	& \multicolumn{1}{c}{IPC}
	& \multicolumn{1}{c}{A}
	& \multicolumn{1}{c}{F}
	& \multicolumn{1}{c|}{BR} 
	& \multicolumn{1}{c}{Broad} 
	& \multicolumn{1}{c|}{Narrow} 
	& \\
\hline
\multicolumn{8}{l|}{\bf Intermediate-mass stars} & & & \\
 Phase I & \color{black} 0\,\% (0/7)
             & \color{black} 43\,\% (3/7) & \color{black} 43\,\% (3/7)
	     & \color{black} 14\,\% (1/7) & \color{black} 0\,\% (0/7)
	     & \color{black} 0\,\% (0/7) 
	     &  \color{black}0\,\% (0/7)
	     &  \color{black}43\,\% (3/7)
	     &  \color{black}0\,\% (0/7)
	     & This study\textcolor{black}{, Ed06} \\

 Phase II & \color{black} 0\,\% (0/7)
	& \color{black} 29\,\% (2/7) 
	& \color{black} 14\,\% (1/7)
	& \color{black} 14\,\% (1/7)
	& \color{black} 0\,\% (0/7)
	& \color{black} 14\,\% (1/7)
	& \color{black} 29\,\% (2/7)
	&  \color{black}0\,\% (0/7)
	&  \color{black}57\,\% (4/7)
	& This study\textcolor{black}{, Ed06} \\

 Phase III & 0\,\% (0/4) & 0\,\% (0/4)
	 & 0\,\% (0/4)
	& 0\,\% (0/4) & 100\,\% (4/4) & 0\,\% (0/4) & 0\,\% (0/4)
	&  0\,\% (0/4)
	&  0\,\% (0/4)
	& This study \\
HAe
	&  7\,\% (2/28)
	&  {11\,\% (3/28)}
	&  36\,\% (10/28)
	&  32\,\% (9/28)
	&  7\,\% (2/28)
	&  7\,\% (2/28) 
	&  0\,\% (0/28) 
	& 36 (10/28)
	& 0 (0/28)
	&  CJ14 \\
\hline
\multicolumn{8}{l|}{\bf Low-mass stars} & & & \\
 CTTS\textcolor{black}{$^a$}
	&  3\,\% (1/39)
	&  \textcolor{black}{21\,\% (8/39)}
	&  36\,\% (14/39)
	&  13\,\% (5/39)
	&  0\,\% (0/39)
	&  3\,\% (1/39) 
	& 26\,\% (10/39) 
	& 38 (15/39) 
	& 28 (11/39)
	&  Ed06 \\
	
WTTS$^b$
	&  33\,\% (2/6) 
	&  0\,\% (0/6) 
	&  0\,\% (0/6)
	&  0\,\% (0/6)
	&  0\,\% (0/6)
	&  67\,\% (4/6) 
	&  0\,\% (0/6) 
	& 0 (0/6)
	& 0 (0/6)
	&  Ed06 \\
\hline 
\end{tabular}
\end{center}
{{\bf Notes.}\\
$^a$The line profiles for CTTSs are categorized as follows:
{group} E: BP Tau; 
{group} DP: CW Tau, RW Aur A,
HN Tau,  {DS Tau,} 
GK Tau, UZ Tau E, V836 Tau, UX Tau;
{group} PC: DR Tau, AS 353A, DL Tau, HL Tau, DG Tau, DF Tau, DO Tau, GG Tau, GW Ori, 
DE Tau, DD Tau, DQ Tau, TW Hya, XZ Tau; 
{group} IPC: AA Tau, BM And, RW Aur B, LkCa8, DN Tau; 
{group} F: GM Aur; 
{group} BR: DK Tau, HK Tau, UY Aur, YY Ori, CI Tau, CY Tau, FP Tau, GI Tau, UZ
Tau W, SU Aur \\
{$^b$}The line profiles for all WTTSs except
 for V826 Tau and TWA 14 are categorized into group F, while those for
 V826 Tau and TWA 14 are categorized into group E.} \\
{\bf Reference:} \\
CJ14: \citet{Cauley2014}; Ed06: \citet{Edwards2006}
\end{table*}


\color{black}

\end{document}